\documentclass[preprint,12pt,authoryear]{elsarticle}
\usepackage{amssymb}
\usepackage{amsmath}
\usepackage{newunicodechar}
\newunicodechar{−}{\ensuremath{-}}
\usepackage[colorlinks=true,
            linkcolor=blue,
            citecolor=blue,
            urlcolor=blue]{hyperref}
            
\usepackage{lineno}
\usepackage{float}
\usepackage{booktabs}
\usepackage{graphicx}
\usepackage[a4paper, margin=1in]{geometry}
            
\journal{Advances in Space Research}

\begin{document}

\begin{frontmatter}

\title{A Multi-Diagnostic Observational Framework for Magnetosonic Solitary Waves During Geomagnetic Storms in Solar Cycles 24 and 25 using Cluster II Mission} 

\author{Murchana Khusroo\corref{cor1}\fnref{label1}} 
\ead{dr.mkhusroo@gmail.com}
\author{Yimnasangla\fnref{label2}}
\affiliation[label1]{organization={Department of Physics, University of Science and Technology Meghalaya},
            addressline={Baridua}, 
            city={Ri-Bhoi},
            postcode={793101}, 
            state={Meghalaya},
            country={India}}

\affiliation[label2]{organization={School of Physical Sciences, Indian Institute of Technology},
             city={Mandi},
             postcode={175075},
             state={Himachal Pradesh},
             country={India}}

\begin{abstract}
Solitary structures, commonly known as solitons, are a class of nonlinear plasma waves that are abundantly found in near-Earth plasmas and planetary magnetospheres. They are nonlinear, localized plasma waves that maintain their shape and velocity over time and distance. While their occurrence in various space plasma environments has been extensively reported, their observation during geomagnetic storms, large-scale disturbances driven by interactions between the solar wind and Earth's magnetosphere, remains limited. In this study, we present a comparative investigation of magnetosonic soliton signatures during geomagnetic storms associated with Solar Cycles 24 and 25. Using high-resolution in-situ magnetic field measurements from the Cluster II mission, we systematically examine the plasma conditions favorable for soliton generation and their evolution during storm-time dynamics. A comprehensive multi-diagnostic observational framework, incorporating several state-of-the-art analytical techniques, is developed to reliably detect and characterize magnetosonic solitons. The results demonstrate that solitary structures in both storms predominantly occur during the early storm intervals, prior to the main phase, suggesting that they may serve as potential precursor signatures of enhanced geomagnetic activity.
\end{abstract}


\end{frontmatter}

\section{Introduction}
\label{sec1}

Space plasmas, which constitute more than 99\% of the visible universe, exhibit rich collective behavior arising from the interaction of charged particles with electromagnetic fields. Understanding the fundamental processes that govern these plasmas is essential for explaining a wide range of space phenomena. Among the most intriguing features of space plasma dynamics are localized, coherent nonlinear structures known as solitary waves or solitons. These structures emerge from a delicate balance between wave dispersion and nonlinear steepening, allowing them to maintain stability and localization as they propagate through a plasma environment \citep{Chen1984, Dauxois2006}.

Solitary structures have therefore become a central topic in contemporary space plasma research \citep{Khusroo2015,Pickett2021}. Their unique stability and localization make them key diagnostic tools for probing plasma conditions in diverse regions of near-Earth space, including the magnetosphere \citep{Lakhina2021, Stasiewicz2003}. Owing to their non-dispersive nature and capacity to transport energy over large distances, solitons have become a subject of sustained interest in space and plasma physics, particularly in contexts involving energy transfer and particle acceleration \citep{Stasiewicz2004, Stasiewicz2006}. These structures have been widely reported in planetary magnetospheres, especially within Earth's magnetosphere and cometary environments \citep{Moolla2003, Matsumoto1994}.  A detailed description of these nonlinear plasma waves/phenomena is discussed below.

\subsection{Solitary Structures}
\label{subsec1}

Solitary structures, often referred to as solitons or solitary waves, are localized and stable single wave pulses that can propagate long distances while maintaining their shape. Their stability arises from a balance between nonlinear and dispersive effects: nonlinearity enhances or steepens the wave amplitude, while dispersion tends to spread it, resulting in a self-reinforcing structure that maintains its form during propagation \citep{Dauxois2006}. In space plasmas, such structures typically form through complex interactions between the solar wind and the terrestrial magnetosphere \citep{Chen1984,Stasiewicz2004,Stasiewicz2003,Trines2007, Khusroo2015, Bora2019}. 

Solitary waves or solitons can be either electrostatic or electromagnetic, depending on the nature of the linear plasma wave mode responsible for perturbing the medium. Electrostatic Solitary Waves (ESWs) are localized pulses sustained by electric fields, often generated by kinetic electron or ion beams, and are characterized by a short-duration, bipolar electric field spike (typically lasting milliseconds) with little to no corresponding magnetic perturbation. These are pulses of pure electric energy and are often observed in active, thin layers of plasma, in dynamic boundary layers such as the auroral acceleration region, the plasma sheet boundary layer, and the magnetosheath, where field-aligned currents and kinetic streaming instabilities are common \citep{Khusroo2015, Shamir2025, Pickett2021, Matsumoto1994, Williams2006}. Extending beyond near-Earth environments, ESWs are also recognized in astrophysical plasmas, where they can contribute to energy redistribution and particle equilibration processes, for example, within the weakly ionized “dead zones” of protoplanetary disks \citep{Das2025}.

Electromagnetic solitons, on the other hand, are localized, non-dispersive wave structures that arise from electromagnetic perturbations generated by plasma waves and propagate stably through a plasma due to a balance between nonlinear and dispersive effects. Various forms of electromagnetic solitons can arise depending on the underlying plasma mode \citep{Stasiewicz2005}; however, in this study, we focus exclusively on magnetosonic (MS) solitons. These structures manifest as nonlinear compressive disturbances resulting from the coupled evolution of the magnetic field and plasma density. Magnetosonic solitary waves (or magnetic pulses) are characterized by localized humps or dips in the magnetic field magnitude $|B|$, where the magnetic pressure acts as the dominant restoring force. They represent pulses that effectively squeeze and release the magnetic field, embodying a hallmark of nonlinear magnetohydrodynamic (MHD) behavior. Such structures emerge from the balance between nonlinear steepening and dispersive effects in a magnetized plasma and are commonly observed in regions with strong or turbulent magnetic fields, including the magnetosheath, magnetopause, and interplanetary shocks \citep{Stasiewicz2003, Bora2019, Acharya2022, Cattaneo1998, Pickett2003}. Recent theoretical work by \cite{Stasiewicz2004} demonstrated that pressure anisotropy plays a critical role in their formation, with spacecraft observations revealing these structures as pulse-like magnetic field fluctuations. Building on this framework, \cite{Bora2019} showed that finite Larmor radius (FLR) effects significantly influence the formation of aperiodic magnetosonic solitary waves, leading to either increasing or decreasing amplitude profiles. Theoretical studies have demonstrated that large-amplitude magnetosonic waves in high-$\beta$ space plasmas can form stable solitary structures in the form of either magnetic enhancements or depressions (bright or dark solitons), depending on the underlying ion velocity distribution and the sign of wave dispersion, with non-Maxwellian equilibria such as ring or loss-cone distributions playing a critical role in determining the polarity and morphology of the resulting solitary waves \citep{Pokhotelov2007}. 

In this study, we investigate these MS solitons in the geomagnetic storm environment across solar cycles 24 \& 25 \citep{SinghPatel2021, UptonHathaway2023} by utilizing high-resolution magnetic field measurements from the Cluster II mission \citep{Escoubet2001, balogh2001cluster_fgi}, the detailed descriptions of which are provided in the following subsections.

\subsection{Geomagnetic Storms}
\label{subsec2}
Geomagnetic storms (GMSs) are temporary disturbances that occur in the planet's magnetosphere when a large-scale transient solar plasma, such as a strong Coronal Mass Ejection (CME) \citep{Webb2012} or a corotating interaction region (CIR) \citep{Heber1999}, interacts with the planet's magnetic field \citep{ChapmanFerraro1930, Srivastava2009, Lakhina2016, Choi2022}. A transfer of increased energy in the magnetosphere occurs during the interaction, causing an increase in plasma movement and electric current in the magnetosphere, which leads to several phenomena and space weather occurrences \citep{keith2020earth, perreault1978geomagnetic}. The main drivers of GMSs are high-speed streams of solar wind, large bursts of CMEs and indirect causes of solar flares and magnetic reconnection \citep{Lakhina2021, yermolaev2005statistical, Paouris2025}.

These storms are characterized by special measurements called the Dst (Disturbance Storm Time) index, which quantifies the strength of the Earth's ring current, and the Kp (Planetarische Kennziffer) index, which shows the global measure of geomagnetic activity \citep{NOAA_SWPC_GeomagneticStorms}. The GMS is classified into three phases: initial, main, and recovery \citep{perreault1978geomagnetic, Saiz2013}. 
\begin{itemize}
    \item Initial: Also known as Sudden Storm Commencement (SSC). This indicates the arrival of the GMS but is not a universal aspect of the storm. In Dst, it is identified by the sudden increase in the index, increasing from 20 nT to 50 nT in just a few minutes \citep{Afolabi2024}.
    \item Main: It is the most intense part of the storm and lasts up to 2-8 hours. It is denoted by the decrease of Dst to less than -50 nT and can go as low as -600 nT \citep{kwak2024g5storm}. 
    \item Recovery: It is the longest phase identified when the Dst levels return to their quiet time values \citep{Mishra2024, Sarma2024}.
\end{itemize}
The Kp indices measure the strength of the storm. On a scale of 0-9, with 0 being the calmest and 9 being the strongest. 0-4 denotes quiet to active but not a storm, while 5 and higher denotes a storm \citep{miyashita2023magnetospheric}. 

\subsection{Cluster II mission}
\label{subsec3}
The Cluster mission was a space mission of the ESA in collaboration with NASA to study the Earth's magnetosphere and near-space environments in three dimensions \citep{Escoubet2001}. The mission was composed of four spacecraft moving in a tetrahedral formation, each carrying 11 instruments. In this study, the instruments used are FGM (Fluxgate magnetometer) and CIS (Cluster Ion Spectroscopy). The FGM is used to perform multipoint measurements to study the magnetic phenomena in three dimensions, while the CIS, consisting of two other instruments-- Composition and Distribution Function (CODIF) analyzer and Hot Ion Analyzer (HIA) is used to study the dynamics of the plasma by providing 3D ion distributions and mass per charge composition \citep{balogh2001cluster_fgi}. In this study, the data from spacecraft C1(Rumba) showed more prominent data compared to the other spacecraft; therefore, it has been used as our base. The mission was initially set to operate for two years but has been in service for almost 20 years, ending its operation in 2024 \citep{masson2024_pioneer_cluster}. The Cluster Science Archive (CSA) \citep{ESA_CSA_Archive} has played a pivotal role in advancing magnetospheric and space plasma research by providing comprehensive access to the full scientific data set of the Cluster II mission, spanning more than two decades. The archive hosts well-calibrated magnetic field, plasma, particle, and wave measurements together with orbit and auxiliary products, enabling detailed investigations of magnetospheric structure and dynamics.  Its consistent data formats and comprehensive metadata have made it especially valuable for studies of multi-scale processes such as boundary layer physics, solar wind–magnetosphere coupling, and kinetic plasma phenomena \citep{Dunlop1995, Paschmann1998}. By ensuring long-term data preservation, reproducibility, and interoperability, the CSA has become an essential resource for both observational studies and theory-driven analyzes, supporting a wide range of investigations from large-scale geomagnetic storm dynamics to localized nonlinear structures in near-Earth space.

Despite extensive observational and theoretical investigations, the role of nonlinear solitary structures in the context of geomagnetic storms and solar cycle variation remains poorly understood. In particular, it is not yet clear whether MS solitary waves systematically occur prior to the onset of geomagnetic storms or whether they can serve as early indicators of impending large-scale magnetic activity. While these nonlinear waves have been widely reported in magnetospheric plasmas, especially in boundary regions such as the magnetopause and magnetosheath \citep{Stasiewicz2003, Stasiewicz2004, Bora2019}, their occurrence in broader space plasma environments, including the solar wind, remains sparsely explored. Observational evidence of MS solitons in heliospheric plasmas is especially limited, primarily due to the transient, localized, and aperiodic nature of these structures, which makes their detection challenging using conventional analysis techniques.

Motivated by these gaps, the present study aims to investigate whether MS solitary structures preferentially emerge during the early phases of geomagnetic storms and whether their occurrence may reflect a characteristic nonlinear response of the plasma to enhanced solar wind driving prior to the main-phase development of the storm. To address this question, we develop and implement a multi-diagnostic observational framework designed specifically to identify localized nonlinear MS waves in space plasmas. The framework combines the current state-of-the-art techniques, which are discussed in the following sections. This integrated approach is intentionally designed to be robust against limited temporal resolution and background turbulence, enabling the reliable detection of solitary structures even in spacecraft data with relatively low cadence. Although solitary waves are recognized as efficient mediators of localized energy transport in collisionless plasmas \citep{Stasiewicz2006}, it remains unclear whether their occurrence and characteristics evolve systematically with storm intensity and across different phases of solar activity \citep{watari2017cycle24, Paouris2025}. The novelty of this work lies in the focused investigation of MS solitary structures through a comparative observational analysis across the geomagnetic storms of Solar Cycles 24 \& 25 \citep{SinghPatel2021, UptonHathaway2023}, using high-resolution measurements from the Cluster II mission. 

In this study, we extend recent advances in solitary structure investigations to systematically examine their properties and variability across different geomagnetic and solar cycle conditions, with a primary focus on the near-Earth space environment and magnetosphere \citep{keith2020earth}. These regions, characterized by strong plasma gradients, boundary layer interactions, and kinetic-scale processes, provide favorable conditions for nonlinear wave generation. Through a comparative analysis of solitary structures observed during Solar Cycles 24 \& 25, we investigate variations in their occurrence rates, morphology, and dynamical behavior \citep{Kumar2022}. This comparison aims to clarify the role of solitary structures in magnetospheric energy transport and their influence on geomagnetic storm evolution under varying solar activity levels \citep{miyashita2023magnetospheric}. Furthermore, by developing a systematic methodology for detecting MS solitons beyond conventional magnetospheric observations, this study establishes a framework for identifying nonlinear solitary wave activity across broader space plasma environments, including solar wind, coronal plasma, heliospheric shock, and sheath regions \citep{RICHARDSON2011} associated with interplanetary disturbances as measured by current and future space missions. Finally, this work evaluates the potential of MS solitons as precursor-like signatures of enhanced magnetospheric activity, thereby advancing our understanding of nonlinear plasma processes associated with the onset and evolution of geomagnetic storms.

\section{Methodology}
\label{sec2}
This study employs an observational analysis of high-resolution magnetic field data from the Cluster II mission to detect and characterize solitary wave structures during GMS. The methodology comprises data selection, preprocessing, a multi-stage detection technique, and advanced signal analysis, which ensure reliable identification and physical characterization of solitary structures.

\subsection{Data Selection and Preprocessing}
\label{subsec2.1}
We selected major GMSs from Solar Cycles (SCs) 24 (\ref{subsec_SC24:Data Selection and Preprocessin SC24})and 25 (\ref{subsec_SC25:Data}) based on their intensity and the availability of full-resolution Cluster data. The standout events chosen for detailed analysis were the March 15-17, 2015 storm (SC24) \citep{Wu2016_StPatrickStorm} and the April 23–24, 2023 storm (SC25) \citep{Paouris2025, kwak2024g5storm}. For each event, full-resolution (22/67 Hz) magnetic field vector data were retrieved from the FGM instrument onboard Cluster spacecraft 1 (C1) via the Cluster Science Archive (CSA) \citep{Escoubet2001, balogh2001cluster_fgi}. Geomagnetic indices (Dst \& Kp) were sourced from the World Data Center for Geomagnetism, Kyoto \citep{WDC_Kyoto} and the GFZ Helmholtz Center for Geosciences \citep{GFZ_KpIndex} to define the different phases of each storm \citep{miyashita2023magnetospheric}.

\subsection{Solitary Structure Detection}
\label{subsec2.2}
The detection process first identified storm phases using the Dst index profile \citep{perreault1978geomagnetic}. Within each phase, the FGM time-series data were systematically scanned in $\sim$10-minute windows across all four spacecraft \citep{Stasiewicz2003} and the best one was identified in C1. Candidate solitary structures were taken based on the visual identification of localized, wave-like pulses or dips in the magnetic field for both SC24 (\ref{fig:solitons_SC24}) and SC25 storm (\ref{fig:gms_solitons_SC25}) events \citep{Pickett2021, Bora2019}. A candidate was confirmed only if an identical, coherent signature was observed in the same time window across multiple spacecraft, verifying its spatial coherence and ruling out instrumental noise \citep{Paschmann1998}.

\subsection{Analysis Techniques}
\label{subsec2.3}
The confirmed candidates were subjected to a suite of analytical techniques \citep{Paschmann1998, Dunlop1995}:
\begin{itemize}
\item \textit{Minimum Variance Analysis (MVA)}: Performed to rotate the magnetic field data into a coordinate frame defined by the structure's principal axes. This clarifies the wave's polarization and propagation geometry. The associated eigenvalues quantify the coherence of the structure \citep{Sonnerup1967, Khrabrov1998, Stasiewicz2004, Bora2019}.
\item \textit{Hodograms}: Plotted in the MVA frame to visualize the magnetic field vector's trajectory, helping to distinguish between different types of solitary structures \citep{Khrabrov1998}.
\item \textit{Continuous Wavelet Transform (CWT)}: Applied to generate time-frequency maps, isolating transient, coherent energy packets characteristic of solitons from background turbulence \citep{Sadowsky1994, Torrence1998, Daubechies1992}. The CWT provides time-localized spectral analysis essential for detecting non-stationary wave packets \citep{Koga2003}. It has also been widely employed to identify soliton and rogue-wave–like structures by revealing their characteristic spatio-temporal localization and broadband spectral signatures arising from nonlinear–dispersive interactions, thereby distinguishing them from linear or stationary wave modes \citep{Pathak2017, Pathak2016}.
\item \textit{Power Spectral Density (PSD) Analysis}: The magnetic field power spectral density was computed using the Welch method to examine the broadband spectral characteristics of the fluctuations across relevant frequency ranges \citep{Welch1967}. Unlike linear or periodic wave modes, which exhibit narrowband spectral peaks, solitary waves are inherently nonlinear and aperiodic, and therefore, produce smooth, power-law–like spectra without distinct spectral lines \citep{Bora2019}. The absence of pronounced spectral peaks, together with the presence of spectral breaks and scale-dependent slopes, provides independent evidence for the nonlinear nature of the observed structures. The PSD analysis complements the time–frequency diagnostics from the CWT by characterizing the global spectral behavior of the fluctuations and identifying dispersive scale transitions associated with kinetic effects \citep{Chen2014}. 
\item \textit{Plasma Parameters}: A comprehensive set of key plasma parameters was calculated and plotted, including Alfvén speed ($v_A$) \citep{Alfven1941}, plasma speed ($v_p)$, ion gyroradius ($\rho_i$), plasma beta ($\beta$), temperature anisotropy ($T_{\perp}/T_{\parallel}$) \citep{Khusroo2015}, Alfvénic Mach number ($M_A$), ion inertial length ($\lambda_i$ ), plasma density and ion plasma frequency ($\omega_{pi}$). These parameters provide essential physical context for interpreting solitary wave signatures. The Alfvén speed and Mach number constrain the wave's propagation velocity relative to characteristic speeds in the plasma. The ion gyroradius and inertial length establish the kinetic scales relevant for dispersive effects that balance nonlinear steepening in solitary waves \citep{Bora2019}. Temperature anisotropy can indicate free energy sources for wave generation. Crucially, plasma beta ($\beta$) confirms a pressure-balanced solitary structure, distinguishing it from a simple compressive pulse or other transient disturbances \citep{Howes2024}.
\end{itemize}
This multi-technique approach allows for the physical identification, classification, and comparative analysis of solitary waves across different storm phases and solar cycles \citep{Trines2007}.

\subsection{Spacecraft Location Mapping}
\label{subsec2.4}
For each confirmed candidate event, the precise location of the Cluster spacecraft was determined using the NASA 4D Orbit Viewer tool. The time references for the events were taken from the FGM data files, and the corresponding spacecraft positions in Geocentric Solar Ecliptic (GSE) coordinates were retrieved from the tool's interface \citep{Bora2019, Sarma2024}. The radial distance from Earth was calculated as 
\[  R=\sqrt{X^2+Y^2+Z^2}\]
This localization confirmed whether the structures were observed in relevant regions such as the magnetopause, magnetosheath, plasma sheet, or outer magnetosphere, providing essential environmental context \citep{Pickett2003, Cattaneo1998}.

\section{Observations}
In this section, we present a detailed analysis of MS solitary waves identified during geomagnetic storms of March 2015 (SC24) and April 2023 (SC25). The following subsections show the Dst and Kp indices for both storms, where the Kp index indicates the magnitude of the storms and the Dst levels show the measurements of the different phases of the geomagnetic storms, which are discussed further in more detail. The magnetic variations, $B_x$, $B_y$, $B_z$, and $|B|$ of the storms have been plotted using the data from the FGM instrument of the Cluster spacecraft. 

\subsection{Solitary Waves for Solar Cycle 24}
\label{subsec_SC24:Data Selection and Preprocessin SC24}
\begin{figure}[htbp]
  \centering
  \includegraphics[width=\columnwidth]{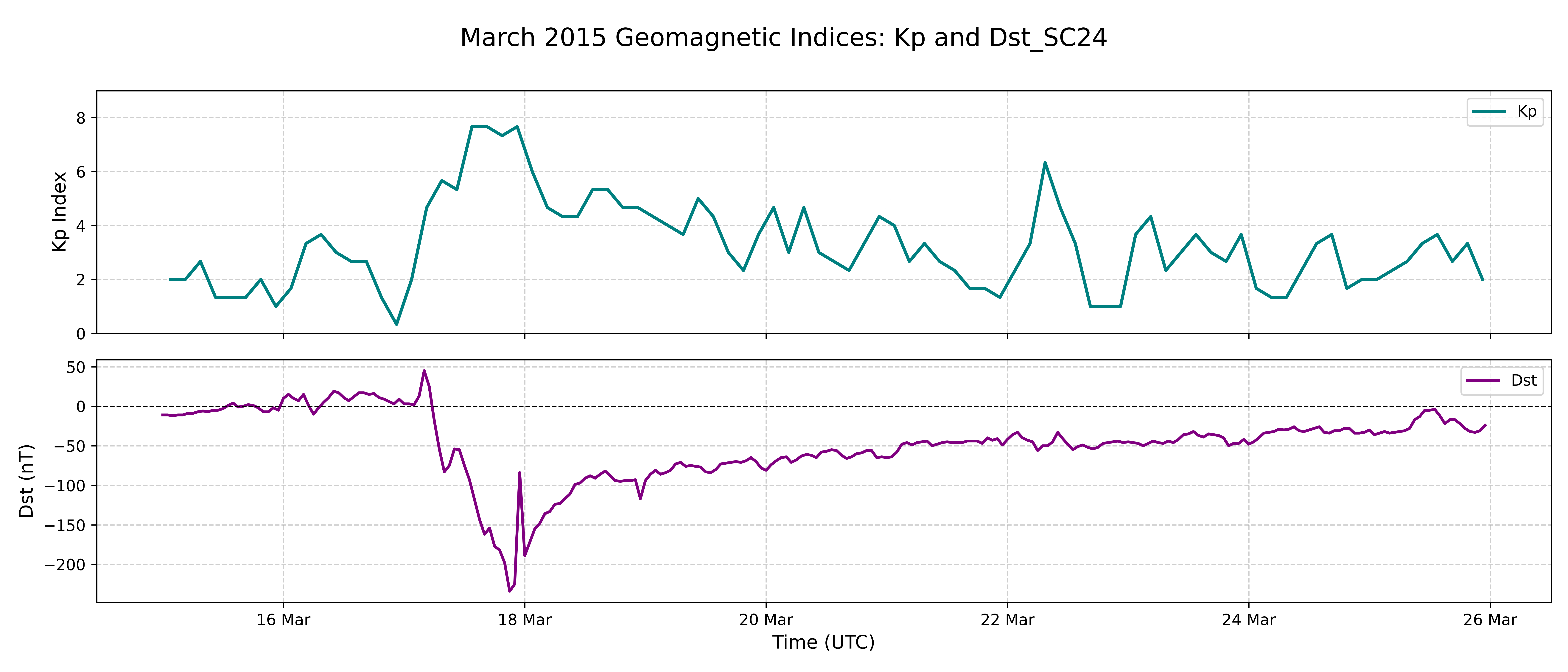}\\
  \caption{Temporal evolution of the geomagnetic indices Kp (top panel) and Dst (bottom panel) during the March 2015 geomagnetic storm associated with SC~24. The storm onset is marked by a rapid enhancement of Kp beginning on 17~March, reaching peak values close to 8, indicative of severe geomagnetic activity. This intensification is followed by a sharp decrease in Dst, which reaches a minimum of approximately $-223$~nT on 17-18~March, corresponding to the main phase of the well-known St.~Patrick’s Day storm. The subsequent gradual recovery of Dst over the following days reflects the decay of the storm-time ring current and the transition into the recovery phase, accompanied by a progressive reduction of Kp to moderate levels.}
  \label{fig:gms_indices_SC24}
\end{figure}

The primary driver of the March 2015 storm, also known as the St. Patrick's Day storm, was recorded to be a halo coronal mass ejection (CME) associated with a strong solar flare (S22W25) and radio bursts that occurred on 15 March, whose interplanetary shock and magnetic cloud hit Earth on 17 March, leading to intense geomagnetic activity when the magnetic field orientation allowed strong coupling with Earth’s magnetosphere. It is classified as a G4 (severe) event using the Kp scale and as a superstorm using the Dst index (Dst $\approx$ −223 nT) \citep{Wu2016_StPatrickStorm}. The Kp index shows a rapid rise to 7–8, indicating strong global disturbance, while the Dst index shows a deep, sharp drop marking a major ring-current–driven storm with a long recovery phase. The Dst and Kp indices are shown in figure~\ref{fig:gms_indices_SC24}.

\begin{figure}[htbp]
  \centering
  \includegraphics[width=\columnwidth]{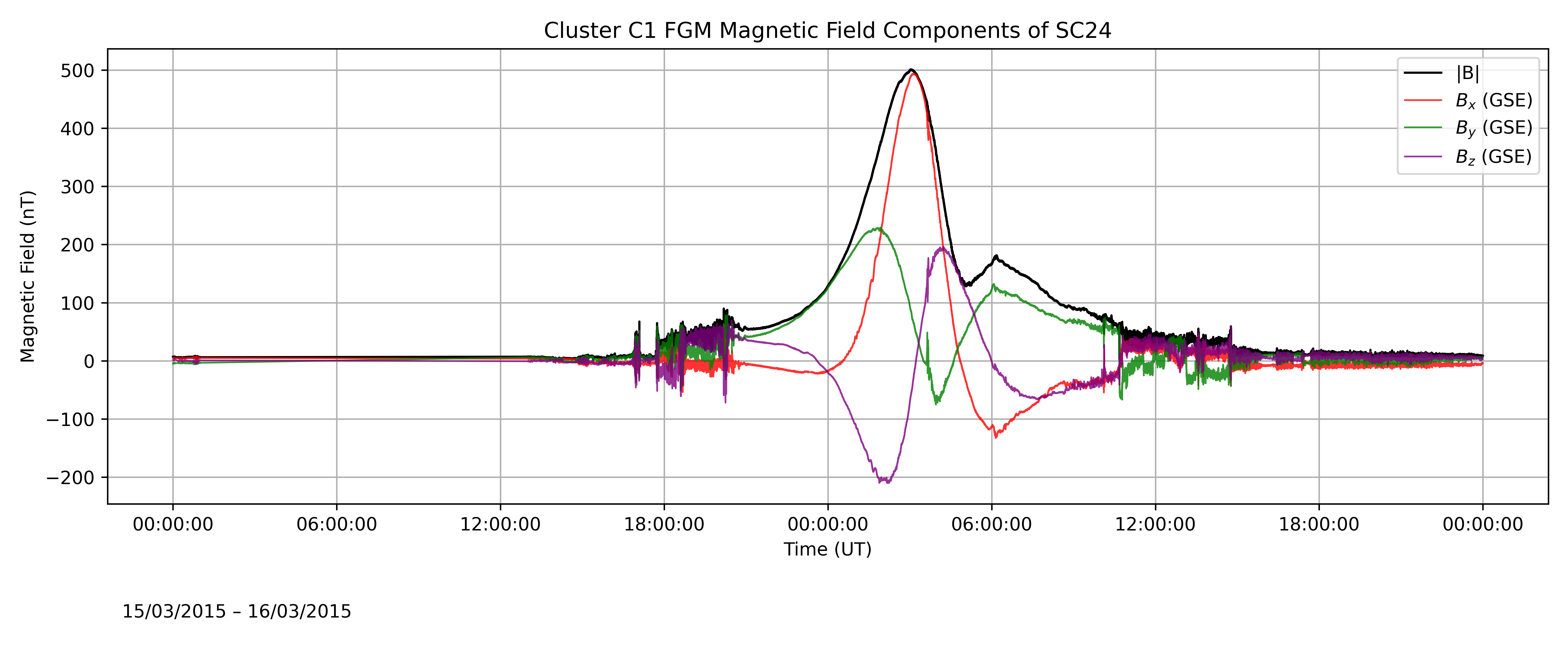}\\
  \caption{Time series of the magnetic field magnitude $|B|$ (black) and the three magnetic field components $B_x$ (red), $B_y$ (green), and $B_z$ (purple) measured by the FGM instrument onboard Cluster~C1, in GSE coordinates during 15-16~March~2015, corresponding to the St.~Patrick’s Day geomagnetic storm (SC~24). The interval captures the storm onset and early main phase, characterized by a strong magnetic field enhancement associated with magnetospheric compression, followed by pronounced, non-sinusoidal fluctuations in the field components. The large-amplitude, asymmetric variations and sharp gradients observed in both $|B|$ and the individual components indicate highly disturbed, nonlinear magnetospheric conditions, providing a favorable environment for the generation of nonlinear plasma structures.}
  \label{fig:gms_magneticfield_SC24}
\end{figure}

The C1 magnetic field measurements from figure~\ref{fig:gms_magneticfield_SC24} show a sharp and coherent increase in magnetic field magnitude, accompanied by large rotations and fluctuations in the individual components around 00:00–06:00 UT on 16 March 2015. This interval aligns precisely with the sudden commencement and the onset of the Dst main phase. The large peak in $|B|$ suggests that the spacecraft was inside a highly compressed magnetospheric region immediately following the CME-driven shock \citep{Astafyeva2015}. The strong negative excursion in $B_z$ coincides with the period when Dst drops rapidly to approximately $\sim$−220 nT, indicating enhanced solar wind–magnetosphere coupling. Thus, the magnetic field signatures recorded by Cluster provide local, high-resolution confirmation of the global geomagnetic storm evolution reflected in the Dst index.

\begin{figure}[htbp]
  \centering
  \includegraphics[width=\columnwidth]{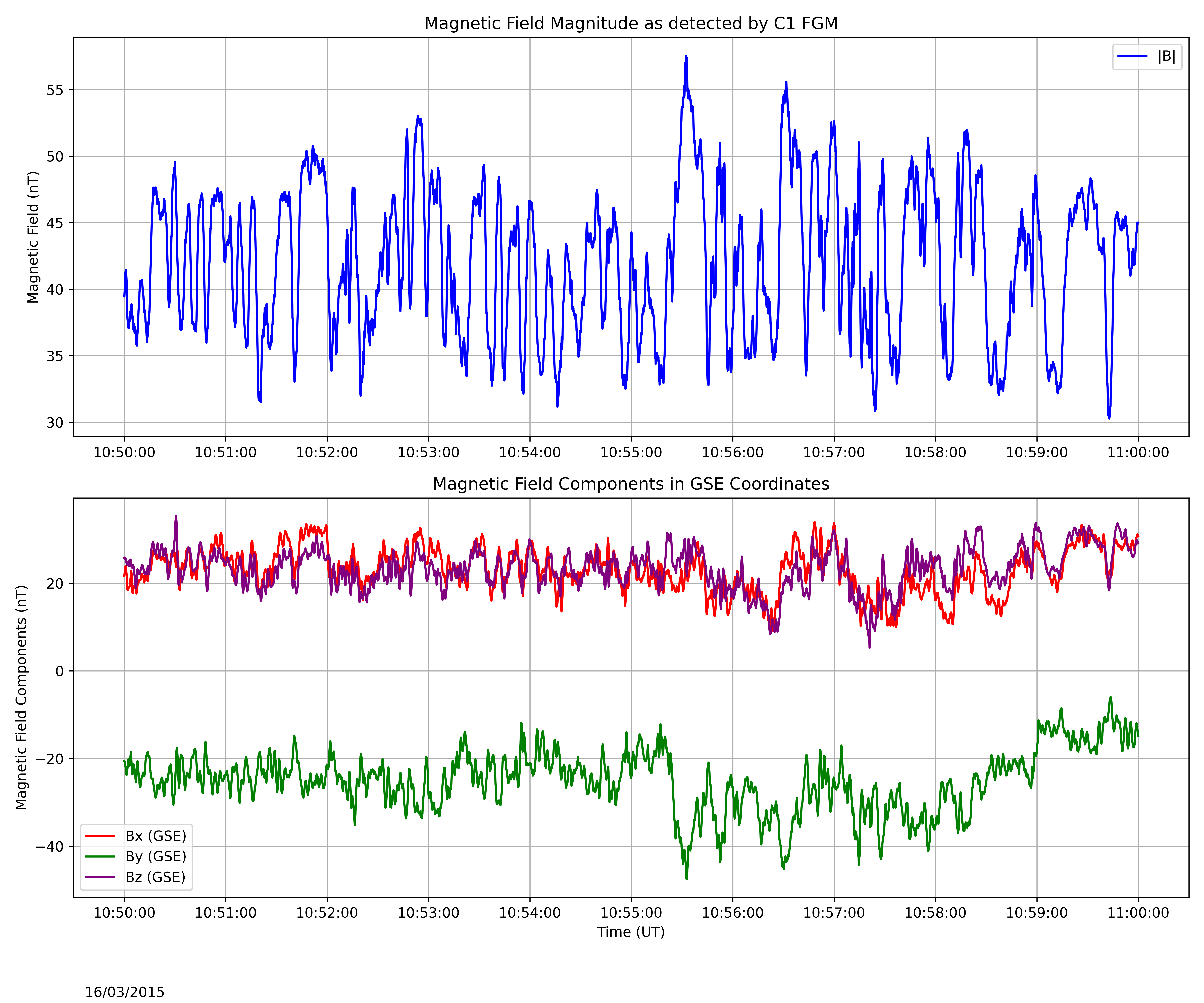}
  \caption{Magnetic field observations from the C1 FGM during the early main phase of the 16 March 2015 geomagnetic storm (Solar Cycle~24). The upper panel shows the magnetic field magnitude, $\lvert B \rvert$, while the lower panel displays the GSE components $B_x$, $B_y$, and $B_z$ over the interval 10{:}50--11{:}00~UT in the magnetopause region.}
  \label{fig:solitons_SC24}
\end{figure}

The FGM data reveal a sequence of small-amplitude, sharp magnetic field fluctuations that are clearly distinguishable from the otherwise relatively quiescent plasma background. These pulses exhibit abrupt onset, short duration, and rapid decay, in contrast to the broader and smoother magnetic variations observed during the subsequent main phase of the storm. A key feature of these structures is their simultaneous appearance of multiple magnetic field components, indicating that they are real, localized coherent structures moving past the spacecraft rather than instrumental noise or random fluctuations. Their isolated nature, well-defined profiles, and rapid rise-and-fall behavior are characteristic of solitary wave phenomena. Accordingly, the early-phase magnetic fluctuations observed on 16 March 2015, between 10:50 and 11:00 UT, are interpreted as solitary structures, as illustrated in figure~\ref{fig:solitons_SC24}. Similar signatures were identified and classified as magnetosonic (MS) solitons by Stasiewicz \citep{Stasiewicz2003, Stasiewicz2004}, while Khusroo \& Bora proved that their aperiodic nature arises from finite Larmor radius (FLR) effects in the plasma \citep{Bora2019}.\\

\begin{figure}[htbp]
  \centering
  \includegraphics[width=\columnwidth]{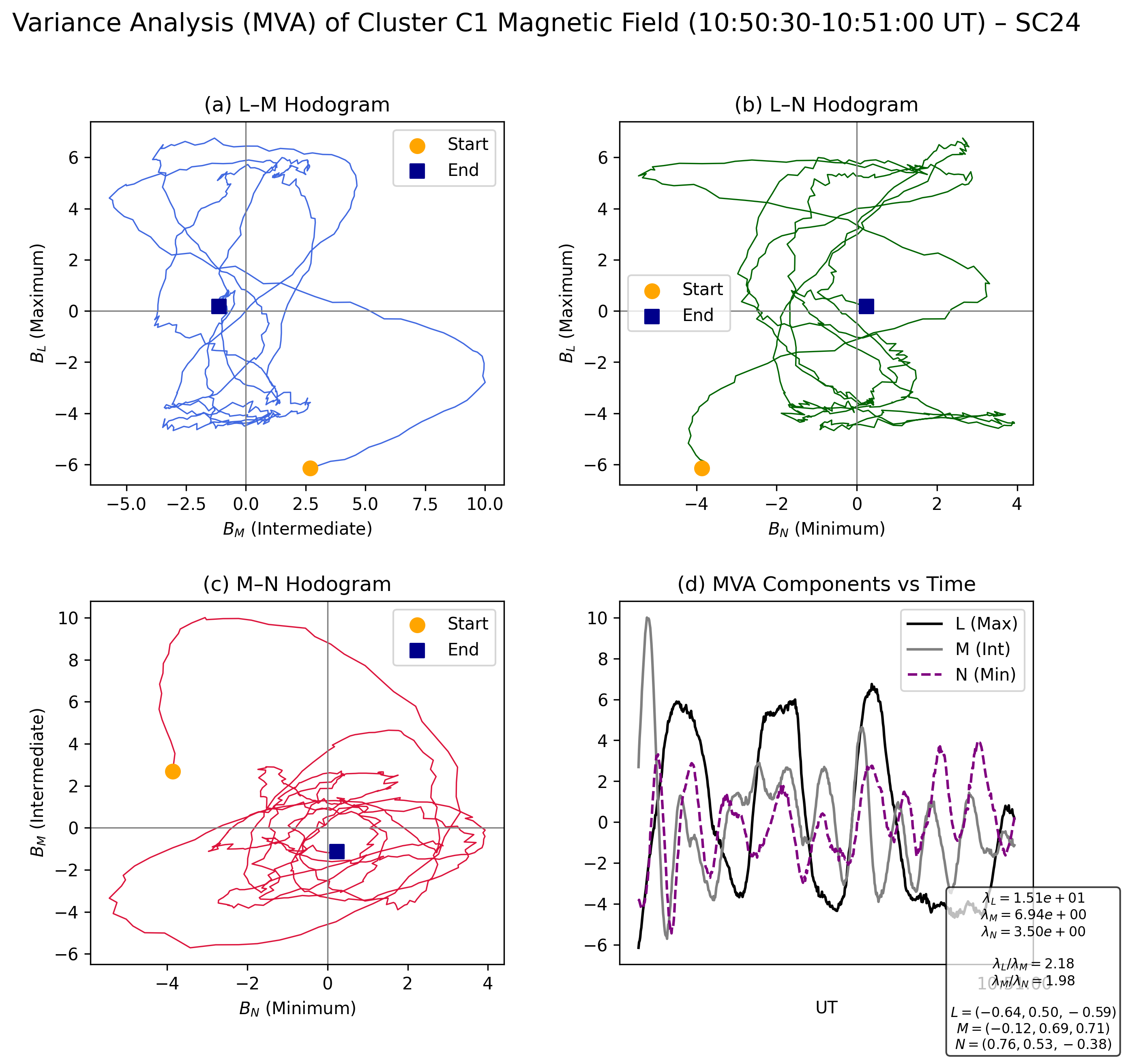}\\
  \caption{Minimum Variance Analysis (MVA) of the magnetic field data from C1 spacecraft during the SC24 interval over a short, localized time window (10:50:30--10:51:00~UT). Panels (a)--(c) show hodograms in the MVA frame corresponding to projections in the $L$--$M$, $L$--$N$ and $M$--$N$ planes, where $L$, $M$, and $N$ denote the maximum, intermediate and minimum variance directions, respectively. The start and end points of each hodogram are marked to illustrate the temporal evolution across the structure. Panel (d) shows the magnetic field components resolved in the MVA frame as a function of time, highlighting the dominance of the maximum variance component and comparatively weaker fluctuations along the minimum variance direction. The corresponding eigenvalues ($\lambda_L$, $\lambda_M$ \& $\lambda_N$), their ratios ($\lambda_L$/$\lambda_M$ \& $\lambda_M$/$\lambda_N$) and eigenvectors (L, M \& N) are indicated, confirming a well-defined variance hierarchy over the selected interval. The short analysis window is chosen to isolate the intrinsic geometry of the soliton train and to minimize contamination from background turbulence.} 
  \label{fig:mva_SC24}
\end{figure}

To examine the geometry of the magnetic structures, we apply minimum variance analysis (MVA) over a short ($\sim$30~s) interval centered on the most prominent solitary features shown in figure~\ref{fig:mva_SC24}. This limited window is chosen to preserve the intrinsic localization of the structures and to avoid contamination from background storm-time fluctuations. The MVA results reveal a clear separation between the maximum (L), intermediate (M) and minimum (N) variance directions, defined by the eigenvectors (L, M \& N) with the corresponding eigenvalues ($\lambda_L$, $\lambda_M$ \& $\lambda_N$) and their ratios ($\lambda_L$/$\lambda_M$ \& $\lambda_M$/$\lambda_N$), indicating a well-defined principal axis of magnetic fluctuations. The hodograms in the L--M, L--N, and M--N planes (panels ac) display compact, bounded, and curved trajectories with structured spread across the intermediate and minimum variance directions \citep{Sonnerup1967, Ma2022}. The absence of elliptical or repetitive patterns rules out linear periodic waves and instead depicts localized nonlinear magnetic structures. Panel~(d) displays the time series of the magnetic field components in the MVA frame. The maximum variance component ($B_L$) dominates the signal and consists of sharp, aperiodic pulses, while the intermediate component ($B_M$) shows weaker but organized fluctuations. The minimum variance component ($B_N$) remains small, indicating that the magnetic field variations are largely confined to a plane. This clear amplitude ordering is consistent with the eigenvalue hierarchy and confirms the presence of a coherent, localized structure. The observed pulse-like behavior in $B_L$, combined with the suppression of $B_N$, is characteristic of nonlinear magnetosonic solitary waves rather than linear or turbulent fluctuations. The sharp, aperiodic pulses observed in the magnetic field time series after performing MVA provide strong supporting evidence for magnetosonic soliton behavior during this interval \citep{Stasiewicz2004,Bora2019}. It is important to note that the amplitudes of the MVA components in panel~(d) do not correspond directly to the absolute magnetic field magnitude shown in figure~\ref{fig:solitons_SC24}, since the MVA is performed on the mean-subtracted magnetic field and represents projections of the fluctuating field onto the principal variance directions. Consequently, the MVA time series highlight the relative organization and coherence of the fluctuations rather than their absolute amplitudes. Despite this difference in scale, the timing of the extrema in the $L$ component closely coincides with the solitary peaks identified in $|B|$, confirming that the soliton pulses dominate the maximum variance direction.

\begin{figure}[htbp]
  \centering
  \includegraphics[width=\columnwidth]{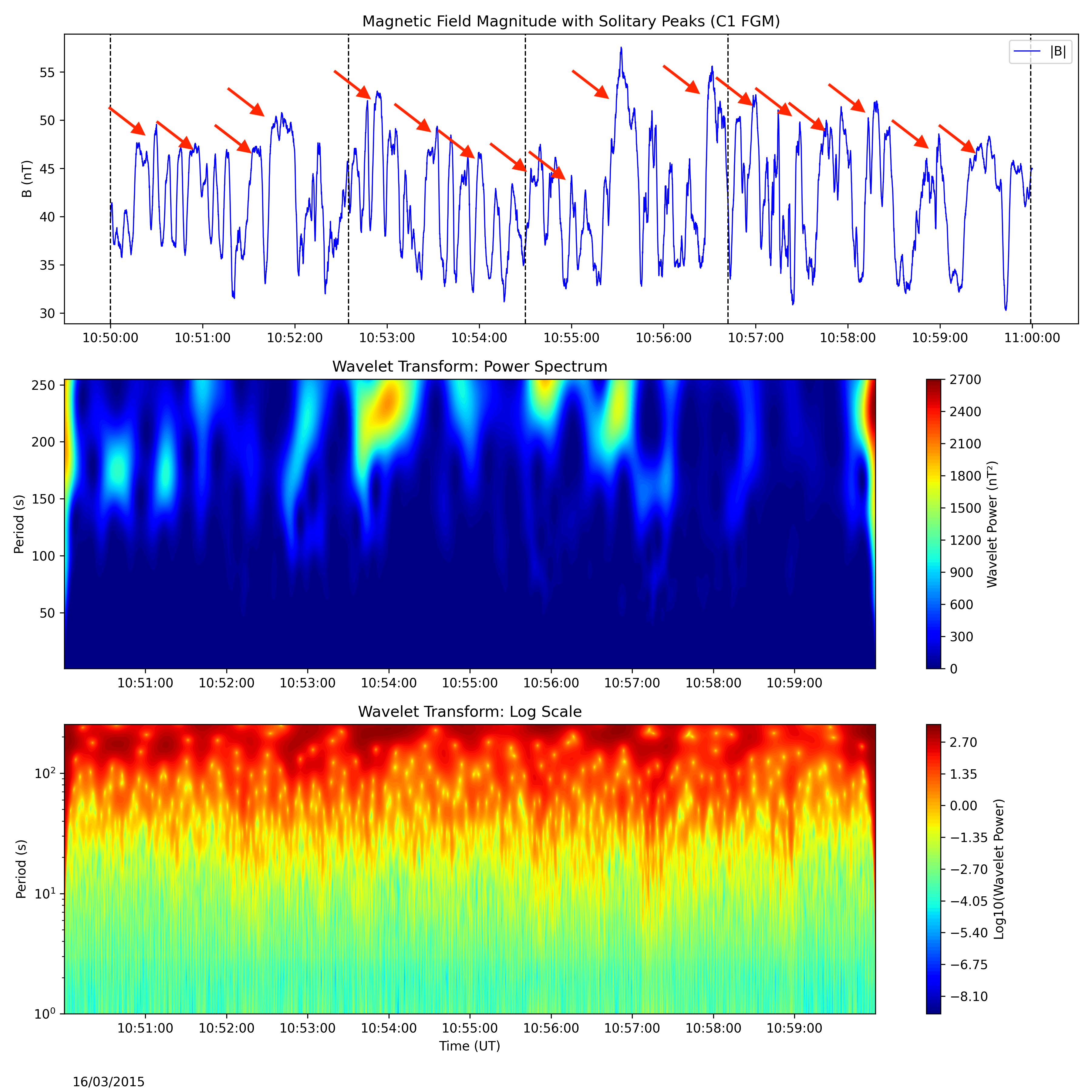}\\
  \caption{Detection of magnetic field solitary structures during SC25 observed by Cluster C1 FGM during a geomagnetic storm interval on 16 March 2015 between 10:50 and 11:00~UT. The top panel shows the magnetic field magnitude $|B|$, where sharp, localized magnetic enhancements marked by red arrows indicate individual solitary pulses. Vertical dashed lines distinguishes successive groups of pulses, suggesting the occurrence of multiple distinct solitary wave trains with irregular spacing and varying amplitudes. The middle panel presents the continuous wavelet transform (CWT) power spectrum of $|B|$, revealing intermittent and temporally localized energy enhancements primarily within characteristic periods of approximately $150$--$250$~s. The color scale represents wavelet power in units of $\mathrm{nT}^2$, with higher intensities (yellow to red; up to $\sim 2.7 \times 10^{3}\ \mathrm{nT}^2$) corresponding to strong magnetic fluctuations associated with the detected structures while the bottom panel displays the logarithmic wavelet power spectrum.}
  \label{fig:wavelet_analysis_SC24}
\end{figure}

Figure~\ref{fig:wavelet_analysis_SC24} illustrates the detection and characterization of magnetosonic solitary structures observed by the Cluster~C1 FGM instrument during the early phase of the March~2015 geomagnetic storm (SC24), between 10:50 and 11:00~UT. The top panel shows the magnetic field magnitude $|B|$ as a function of time, where a sequence of sharp, localized enhancements is clearly visible above the background field of $\sim$35–40~nT. These features, highlighted by arrows, occur intermittently as a train of discrete pulses with peak amplitudes reaching $\sim$50–57~nT. Each pulse exhibits a rapid rise followed by a similarly rapid decay, distinguishing them from the broader, smoother magnetic variations associated with large-scale storm-time compressions and indicating the passage of spatially localized magnetic structures. The middle panel presents the continuous wavelet transform (CWT) power spectrum of $|B|$, which provides a time–frequency representation of the magnetic fluctuations. The color bar on the right denotes the wavelet power in units of nT$^{2}$, with cooler colors (dark blue) corresponding to weak power levels ($\lesssim 10^{2}$~nT$^{2}$) and warmer colors (green to red) indicating progressively stronger power, reaching peak values of $\sim 2.3\times10^{3}$~nT$^{2}$. Each solitary pulse identified in the time series is associated with a localized enhancement of wavelet power, confined to narrow time intervals and spanning characteristic periods from a few tens to a few hundred seconds. Notably, these power enhancements appear as vertically localized patches rather than as continuous horizontal bands, demonstrating that the fluctuations are non-stationary and lack a well-defined single frequency. The bottom panel shows the logarithmic representation of the wavelet power spectrum. In this representation, warm colors indicate enhanced power over a wide range of temporal scales, while cool colors mark background-level activity. The solitary pulses manifest in the wavelet spectrum as vertically elongated regions of enhanced power, appearing precisely at the time intervals corresponding to the pulses identified in the magnetic field magnitude and extending across a range of characteristic periods, indicative of strong temporal localization and multiscale coupling \citep{Pathak2017}. These features are observed during the initial phase of the geomagnetic storm, prior to the main storm impact, suggesting that localized nonlinear processes are already active in the magnetospheric environment before large-scale storm-time reconfiguration becomes dominant. The combined time-domain and time–frequency diagnostics therefore offer a robust framework for distinguishing solitary wave activity from background turbulence and linear wave phenomena.

\begin{figure}[htbp]
  \centering
  \includegraphics[width=\columnwidth]{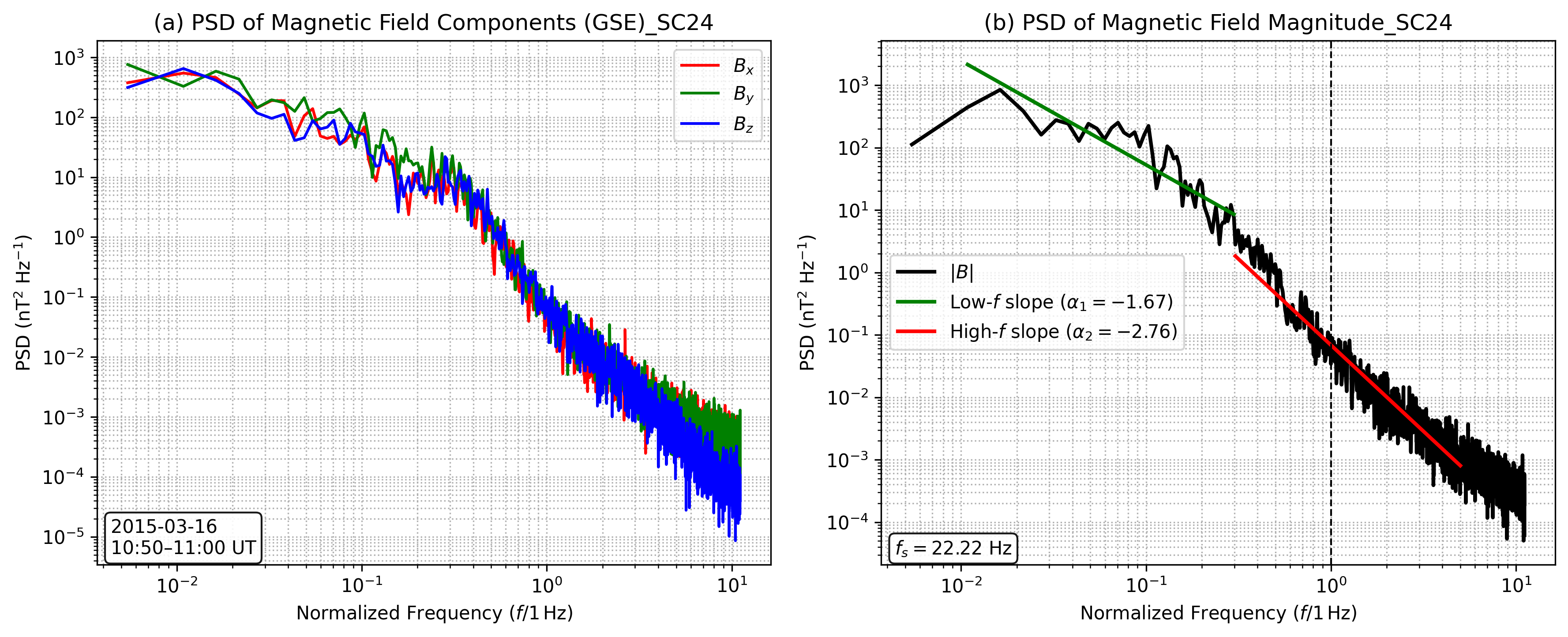}\\
  \caption{Panel (a) shows the power spectral density (PSD) of the magnetic field components ($B_x$, $B_y$, and $B_z$) in GSE coordinates obtained from Cluster~C1 FGM data during 10:50--11:00~UT on 16~March~2015 (SC24). All components exhibit broadband spectra without narrowband peaks, indicating aperiodic and non-monochromatic fluctuations. Panel (b) demonstrates the PSD of the magnetic field magnitude $|B|$, showing a broken power-law behaviour with a spectral break near normalized frequency $f/1\,\mathrm{Hz} \sim 1$ (vertical dashed line). The low-frequency range follows a shallower slope ($\alpha_1 \approx -1.67$), while the high-frequency range steepens to $\alpha_2 \approx -2.76$, consistent with enhanced nonlinear and dispersive effects at smaller scales. The sampling frequency is $f_s = 22.22$~Hz. }
  \label{fig:psd_analysis_SC24}
\end{figure}

Figure~\ref{fig:psd_analysis_SC24} presents the power spectral density (PSD) analysis of the Cluster~C1 magnetic field during 10:50--11:00~UT on 16~March~2015, corresponding to the early phase of the SC24 geomagnetic storm. Panel~(a) shows the PSD of the three magnetic field components ($B_x$, $B_y$, and $B_z$) in GSE coordinates. All components exhibit a continuous, broadband spectral distribution across the resolved frequency range, with no evidence of narrowband spectral peaks. This behavior indicates that the observed fluctuations are not associated with coherent, monochromatic wave modes but instead reflect aperiodic and intermittent dynamics involving coupled variations across multiple magnetic field components. Panel~(b) shows the PSD of the magnetic field magnitude $|B|$, highlighting compressive fluctuations relevant to magnetosonic activity. The $|B|$ spectrum exhibits a clear broken power-law behavior. The vertical dashed line denotes the reference frequency corresponding to the normalization scale ($f/1\,\mathrm{Hz}=1$) and serves as a convenient divider between lower and higher-frequency regimes rather than marking the spectral break itself. The actual spectral steepening occurs at frequencies above this reference scale, indicating that nonlinear energy transfer persists across a broad range of frequencies before dispersive and kinetic effects become dominant. At lower frequencies, the spectrum follows a relatively shallow power law with slope $\alpha_1 \approx -1.67$, characteristic of large-scale, organized magnetic fluctuations. At higher frequencies, the spectrum steepens significantly to $\alpha_2 \approx -2.76$, reflecting the increasing influence of nonlinear steepening balanced by dispersive processes at smaller scales. Crucially, the PSD does not exhibit discrete spectral peaks at any frequency. For linear or weakly nonlinear periodic waves, one would expect well-defined spectral peaks corresponding to a dominant wave frequency and its harmonics \citep{Stoica2005}. The complete absence of such peaks in both the component-wise and $|B|$ spectra provides strong evidence that the fluctuations are intrinsically nonlinear. Instead, the smooth broadband spectrum and broken power-law scaling indicate aperiodic, scale-coupled dynamics consistent with a train of magnetosonic solitary structures rather than a sustained linear wave train. The delayed spectral break at higher normalized frequencies further supports the interpretation that these structures are governed by local nonlinear–dispersive balance rather than by a single characteristic linear timescale, a defining feature of magnetosonic soliton dynamics \citep{Bora2019}.

\begin{figure}[htbp]
  \centering
  \includegraphics[width=\columnwidth]{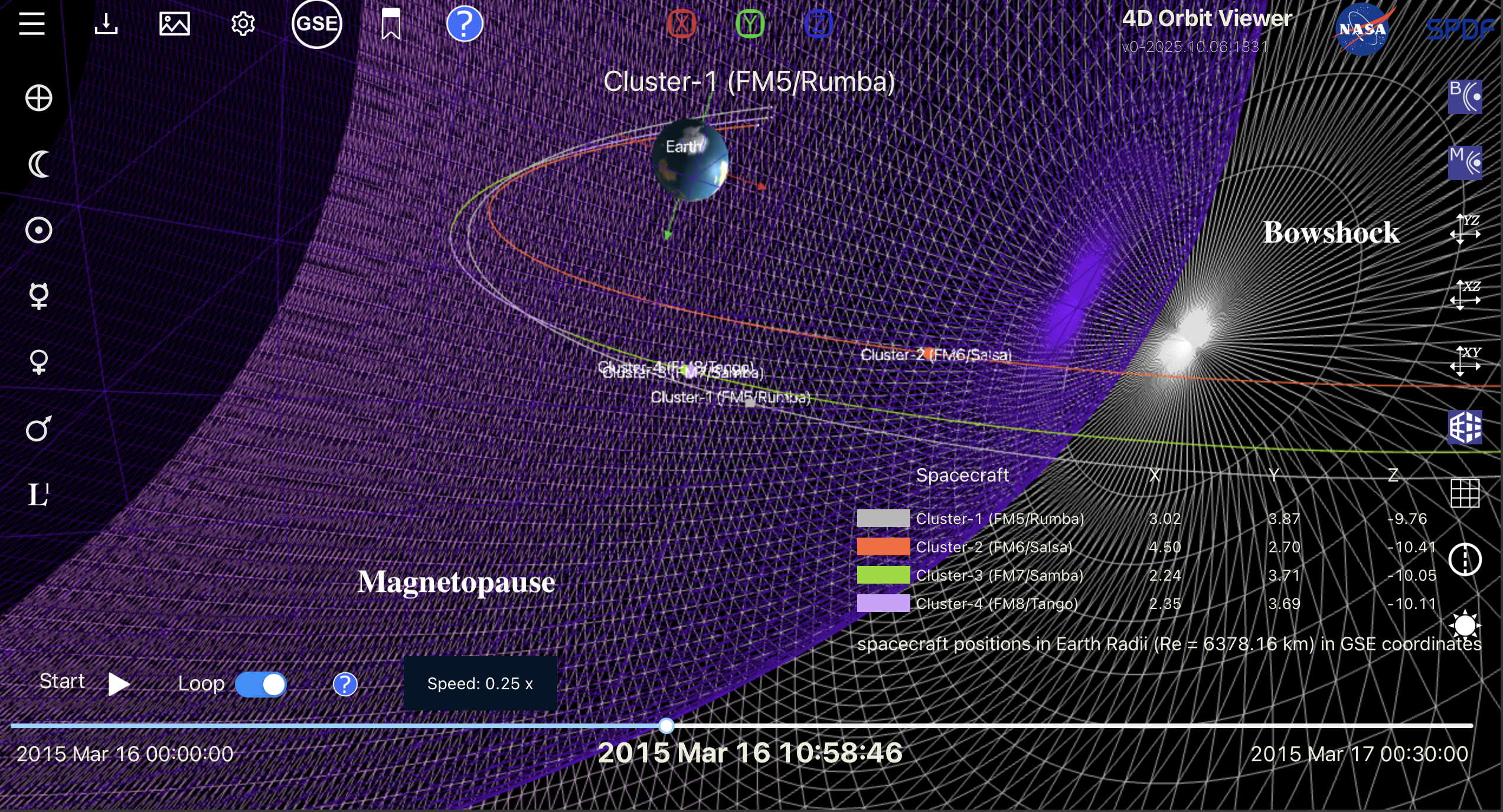}\\
  \caption{Orbital configuration of the Cluster constellation in GSE coordinates during the SC24 geomagnetic storm on 16~March~2015, shown using the 4D Orbit Viewer \citep{ssc4dviewer}. The trajectories of the four Cluster spacecrafts are plotted relative to the modeled bow shock and magnetopause. The snapshot corresponds to $\sim$10:58:46~UT, within the interval 10:50--11:00~UT during which magnetosonic solitary structures were detected by Cluster--1 as it crossed the magnetopause region of the magnetosphere.}
  \label{fig: spacecraft_location_SC24}
\end{figure}

Figure \ref{fig: spacecraft_location_SC24} provides the orbital trajectories of the Cluster spacecraft in GSE coordinates during the SC24 geomagnetic storm on 16~March~2015, shown in 4D Orbit Viewer \citep{ssc4dviewer}. At the time of the solitary wave observations (10:50--11:00~UT), C1 is found to be located in the magnetopause region of the magnetosphere, a plasma environment known to support magnetosonic solitary structures where finite Larmor radius (FLR) effects play a significant role in their formation \citep{Bora2019}. This spatial context reinforces the interpretation that the observed magnetic pulses arise from localized nonlinear processes operating during the early phase of the geomagnetic storm, prior to the main storm-time compression.

\begin{figure}[htbp]
  \centering
  \includegraphics[width=\columnwidth]{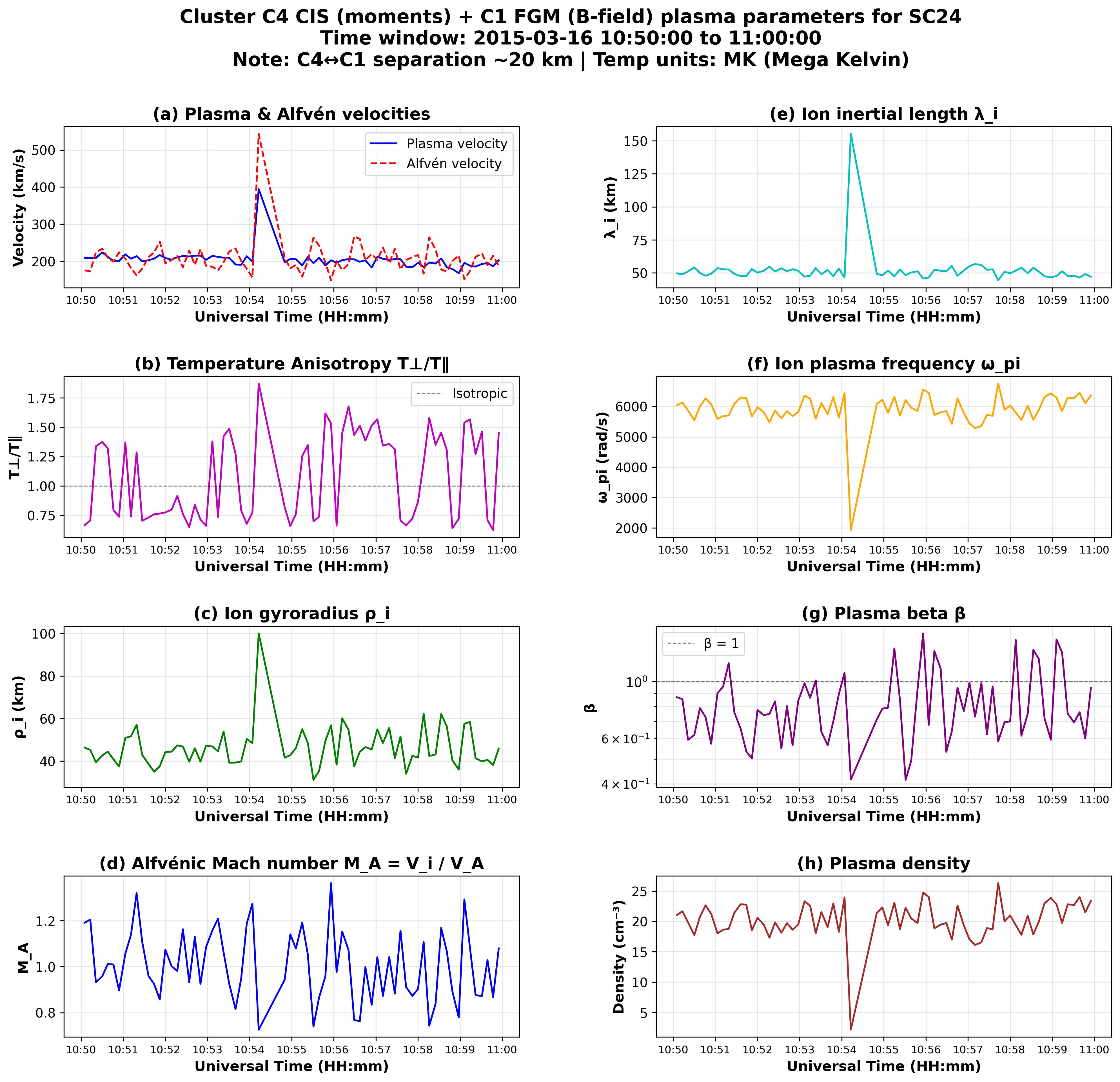}\\
  \caption{The plasma and magnetic field parameters during the interval 10:50--11:00~UT on 16~March~2015 (SC24), derived from Cluster--4 CIS ion moments and Cluster--1 FGM magnetic field measurements. Panels~(a)--(d) display the plasma bulk velocity, Alfv\'en velocity, temperature anisotropy $T_{\perp}/T_{\parallel}$, and Alfv\'enic Mach number $M_A$, respectively. Panels~(e)--(h) show the ion inertial length $\lambda_i$, ion plasma frequency $\omega_{pi}$, plasma beta $\beta$, and plasma density. The C4--C1 spacecraft separation during this interval is $\sim$20~km, ensuring that plasma and magnetic field measurements sample the same localized region. The parameters indicate a low--to--moderate $\beta$ plasma with ion temperature anisotropy, kinetic scale lengths comparable to the observed magnetic structures, and predominantly trans--Alfv\'enic flow conditions, providing a favorable environment for the formation and propagation of nonlinear magnetosonic solitary structures.}
  \label{fig:plasma_parameters_SC24}
\end{figure}

During the interval 10:50–11:00~UT on 16~March~2015, coinciding with the detection of magnetosonic solitary structures, the plasma environment exhibits clear signatures of kinetic–scale and nonlinear dynamics (Figure~\ref{fig:plasma_parameters_SC24}). The plasma bulk velocity remains comparable to the local Alfv\'en speed (Figure~\ref{fig:plasma_parameters_SC24}a), with the Alfv\'enic Mach number fluctuating around unity (Figure~\ref{fig:plasma_parameters_SC24}d), indicating predominantly trans–Alfv\'enic flow conditions rather than strongly super or sub-Alfv\'enic regimes. Such conditions are favorable for the excitation and sustenance of compressive magnetosonic activity. The ion temperature anisotropy (Figure~\ref{fig:plasma_parameters_SC24}b) consistently departs from isotropy, with $T_{\perp}/T_{\parallel} \gtrsim 1$ for much of the interval, indicating the presence of free energy capable of driving nonlinear compressive modes. Simultaneously, the ion gyroradius $\rho_i$ and ion inertial length $\lambda_i$ (Figures~\ref{fig:plasma_parameters_SC24}c and~e) remain on the order of a few tens of kilometers, comparable to the characteristic spatial scales inferred from the observed magnetic pulses. This correspondence suggests that finite Larmor radius and ion inertial effects play a significant role in shaping the observed structures. The plasma beta $\beta$ (Figure~\ref{fig:plasma_parameters_SC24}g) remains close to unity throughout the interval, indicating a near balance between thermal and magnetic pressures, a key requirement for the existence of magnetosonic solitary structures. Variations in plasma density and ion plasma frequency (Figures~\ref{fig:plasma_parameters_SC24}h and~f) are relatively modest and do not exhibit sharp discontinuities, supporting the interpretation that the observed magnetic fluctuations are not associated with shock-like compressions or large-scale discontinuities. However, a notable feature within this interval is the brief but pronounced excursion observed around 10:54–10:55~UT, where several plasma parameters exhibit sharp, transient variations (Figure~\ref{fig:plasma_parameters_SC24}). This interval is characterized by a localized enhancement in both the plasma and Alfv\'en speeds, accompanied by a simultaneous increase in the ion inertial length and ion gyroradius, and a corresponding decrease in plasma density and ion plasma frequency. These variations are mutually consistent and reflect a short-lived, localized density depletion and magnetic field intensification rather than a large-scale shock or boundary crossing. The anti-correlation between $\lambda_i$ and $\omega_{pi}$, together with the density minimum, indicates a transient kinetic-scale structure embedded within the magnetopause boundary layer. Such behavior is characteristic of nonlinear MS solitary structures, which locally modify plasma density, magnetic field strength, and kinetic scales while leaving the surrounding plasma largely undisturbed. The rapid recovery of all parameters to their pre-event levels further supports the interpretation that these excursions arise from localized nonlinear wave activity rather than global magnetospheric reconfiguration. The range of the plasma parameters is shown in Table \ref{fig:Table1_SC24}.

\begin{table*}[htb]
\centering
\caption{Plasma parameters observed during the March 2015 (SC24) }
\label{table:plasma_params}
\begin{tabular}{lp{5cm}p{4cm}p{7cm}}
\toprule
\textbf{S.no} & \textbf{Parameter} & \textbf{Value Range} \\
\midrule
1 & Plasma velocity & 170–400 km/s \\
2 & Alfvén velocity & 130–560 km/s \\
3 & Temperature anisotropy & 0.65–1.85 \\ 
4 & Ion gyroradius & 35–100 km \\
5 & Alfvénic Mach no. & 0.72–1.38 \\
6 & Ion inertial length & 45–155 km  \\
7 & Ion plasma frequency & 2000-6800 rad/s\\ 
8 & Plasma beta & 0.4–1.5 \\
9 & Plasma density & 1-27 cm$^{-3}$\\
\bottomrule
\end{tabular}
\label{fig:Table1_SC24}
\end{table*}

\subsection{Solitary Waves for Solar Cycle 25}
\label{subsec_SC25:Data}
\begin{figure}[htbp]
  \centering
  \includegraphics[width=\columnwidth]{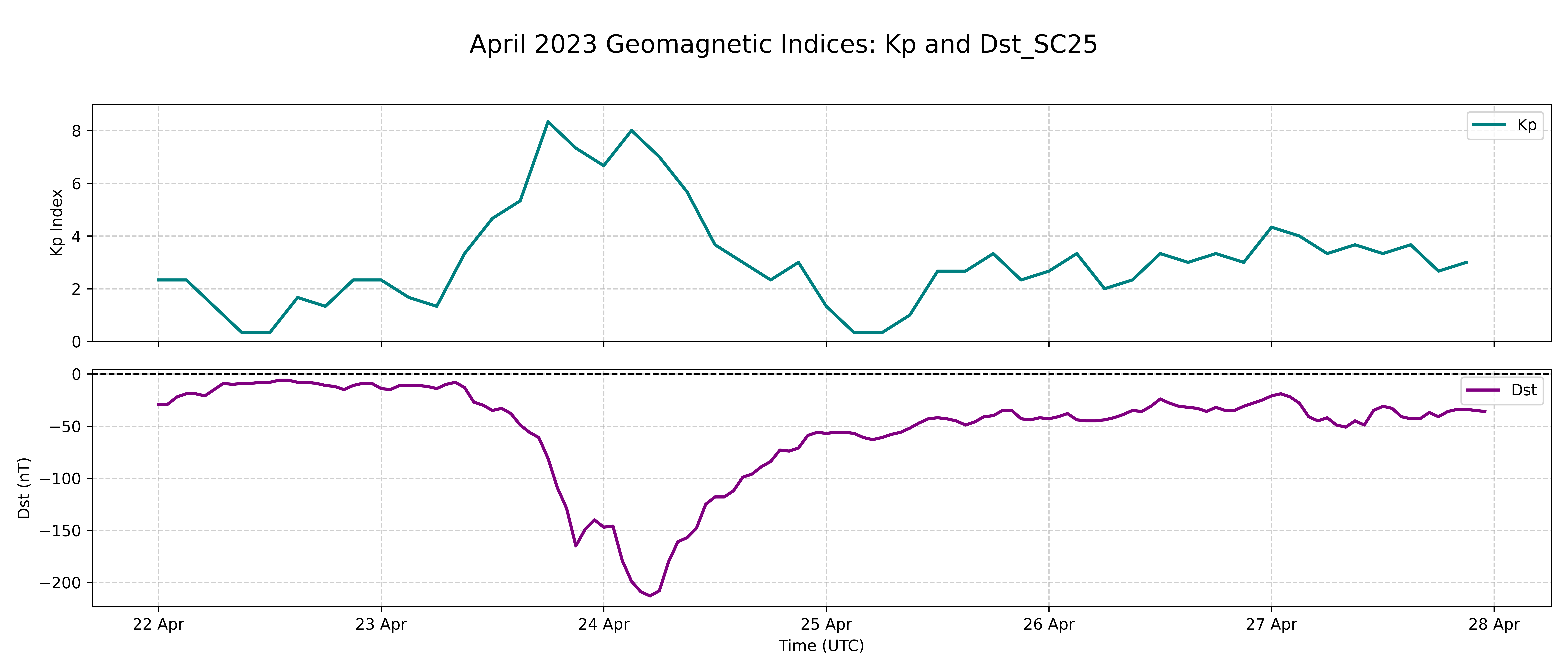}\\
  \caption{Temporal evolution of the geomagnetic indices Kp (top panel) and Dst (bottom panel) during the 22--28~April~2023 geomagnetic storm (Solar Cycle~25). The storm onset is characterized by a rapid intensification of geomagnetic activity, with Kp reaching values above 8, followed by a sharp Dst depression below $-200$~nT on 24~April, indicative of an intense G4-class storm. The subsequent gradual recovery of Dst reflects the decay of the storm-time ring current during the recovery phase.}
  \label{fig:gms_indices_sc25}
\end{figure}

In the case of SC25, the geomagnetic storm was triggered by a single coronal mass ejection (CME) originating from an active region (AR) with relatively weak magnetic field strength, yet it resulted in a severe (G4) geomagnetic response, with the Kp index reaching values close to 8 \citep{Paouris2025}. Figure~\ref{fig:gms_indices_sc25} shows the storm evolution, beginning with the arrival of the interplanetary shock on 23~April, marking a sudden storm commencement and a rapid intensification of geomagnetic activity. During this interval, the Dst index decreases sharply from near-quiet levels ($\sim -10$~nT) to below $-150$~nT, while Kp rises rapidly, indicating strong solar wind-magnetosphere coupling. The main phase of the storm develops late on 23~April and peaks on 24~April, when Dst reaches minimum values of approximately $-210$ to $-230$~nT, concurrent with sustained elevated Kp levels near 8. Subsequently, the storm transitions into a recovery phase characterized by a gradual increase in Dst toward less negative values from 24 to 27~April, accompanied by a progressive reduction in Kp to moderate levels ($\sim$1--3), reflecting the decay of the storm-time ring current and the relaxation of magnetospheric conditions.\\

\begin{figure}[htbp]
  \centering
  \includegraphics[width=\columnwidth]{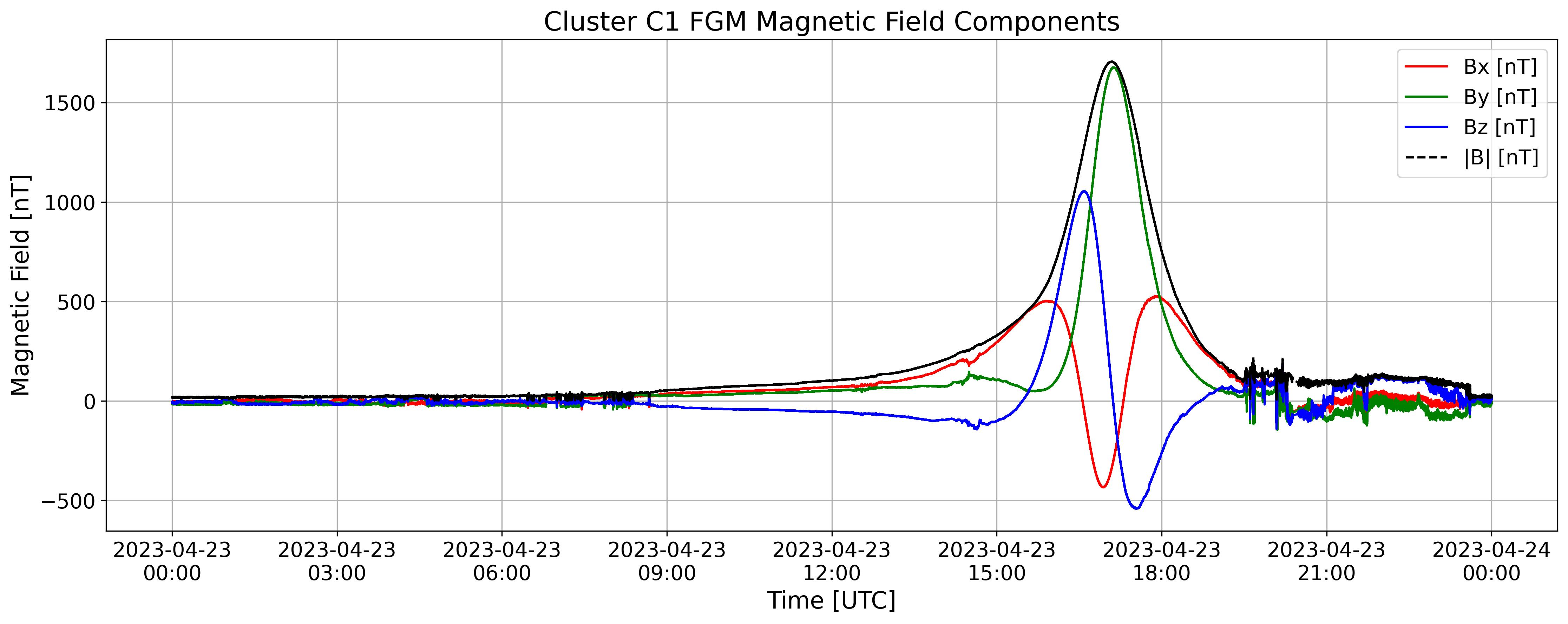}\\
  \caption{Time series of the magnetic field magnitude $|B|$ (black) and the three magnetic field components $B_x$ (red), $B_y$ (green), and $B_z$ (purple) measured by the FGM instrument onboard Cluster~C1, in GSE coordinates during 23-24~April~2023, corresponding to the geomagnetic storm of SC~25. The interval captures the storm onset and early main phase, characterized by a strong magnetic field enhancement associated with magnetospheric compression, followed by pronounced, non-sinusoidal fluctuations in the field components. The large-amplitude, asymmetric variations and sharp gradients observed in both $|B|$ and the individual components indicate highly disturbed, nonlinear magnetospheric conditions, providing a favorable environment for the generation of nonlinear plasma structures.}
  \label{fig:gms_magneticfield_SC25}
\end{figure}

Figure~\ref{fig:gms_magneticfield_SC25} shows the evolution of the magnetic field measured by the Cluster~C1 FGM instrument in GSE coordinates during the SC25 geomagnetic storm on 23--24~April~2023. The figure displays the three magnetic field components ($B_x$, $B_y$, $B_z$) together with the total magnetic field magnitude $|B|$ over a 24-hour interval. During the initial phase of the storm, the magnetic field remains relatively stable with modest fluctuations, indicating a weakly disturbed background prior to the arrival of the storm driver. A pronounced enhancement in $|B|$ is observed near $\sim$17:00--18:00~UT on 23~April, coinciding with sharp, large-amplitude variations in all three field components. The peak magnetic field magnitude reaches values of the order of $\sim$1700~nT, accompanied by strong, rapid excursions in $B_z$, reflecting intense compressional and rotational perturbations of the magnetic field. Following this interval, $|B|$ gradually decreases, and the magnetic field components relax toward quieter levels, marking the transition into the recovery phase of the geomagnetic storm. Superposed on this large-scale storm evolution, shorter-timescale fluctuations are visible, particularly during the storm onset and early main phase, providing the temporal windows in which localized magnetic structures, including solitary wave activity, are subsequently examined in detail.\\

\begin{figure}[htbp]
  \centering
  \includegraphics[width=\columnwidth]{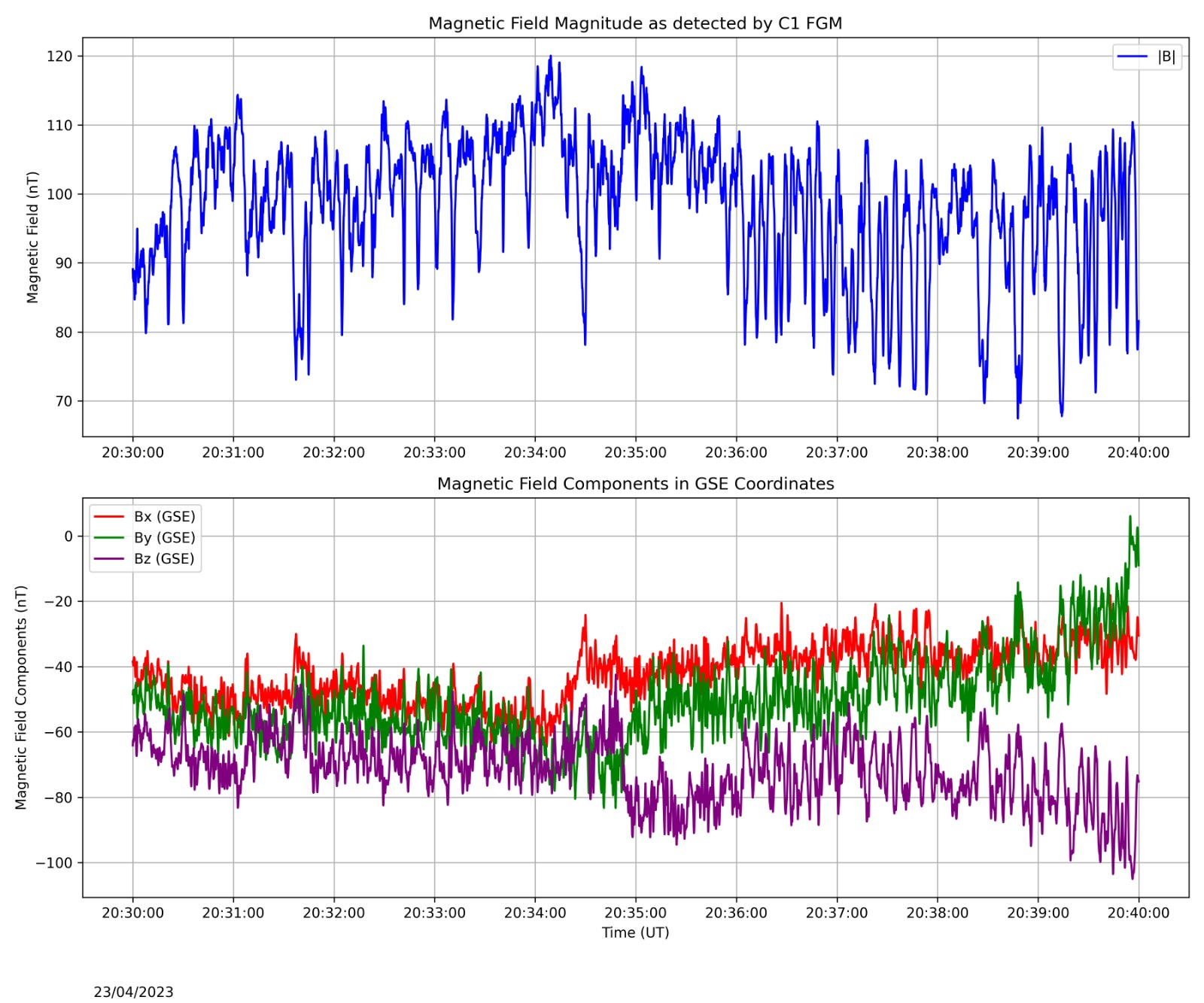}\\
  \caption{Magnetic field observations from the C1 FGM during the early main phase of the 23 April 2023 geomagnetic storm (Solar Cycle~25). The upper panel shows the magnetic field magnitude, $\lvert B \rvert$, while the lower panel displays the GSE components $B_x$, $B_y$, and $B_z$ over the interval 20{:}30--20{:}40~UT in the magnetopause region.}
  \label{fig:gms_solitons_SC25}
\end{figure}

\begin{figure}[htbp]
  \centering
  \includegraphics[width=\columnwidth]{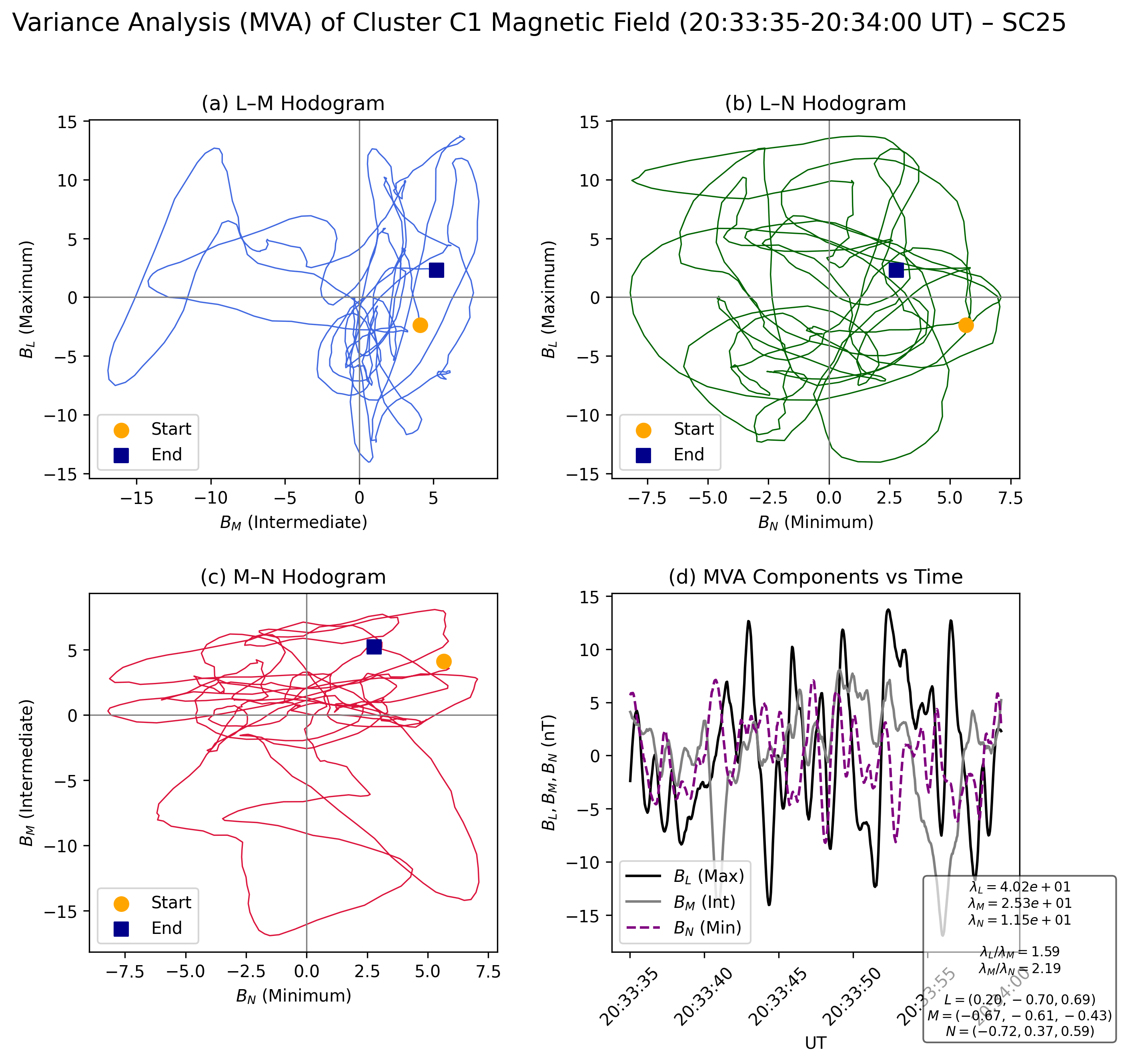}\\
  \caption{Minimum Variance Analysis (MVA) of the Cluster-1 magnetic field during the SC25 interval over a short time window (20:33:35--20:34:00~UT). Panels (a)--(c) display hodograms in the MVA frame corresponding to projections in the $L$--$M$, $L$--$N$ and $M$--$N$ planes, where $L$, $M$, and $N$ represent the maximum, intermediate, and minimum variance directions, respectively. The hodograms show curved, non-elliptical trajectories with finite spread in the intermediate and minimum variance directions, indicating a localized nonlinear magnetic structure rather than a coherent linear wave. Start and end points of the trajectories are marked by circles and squares to illustrate the temporal evolution across the structure. Panel (d) shows the corresponding magnetic field components resolved in the MVA frame as a function of time, highlighting the dominance of the maximum variance component and weaker, phase-shifted variations in the intermediate and minimum directions. The eigenvalues, their ratios, and the corresponding eigenvectors are indicated, confirming a clear but moderate variance separation over the selected interval.}
  \label{fig:MVA_SC25}
\end{figure}

\begin{figure}[H]
\includegraphics[width=\columnwidth]{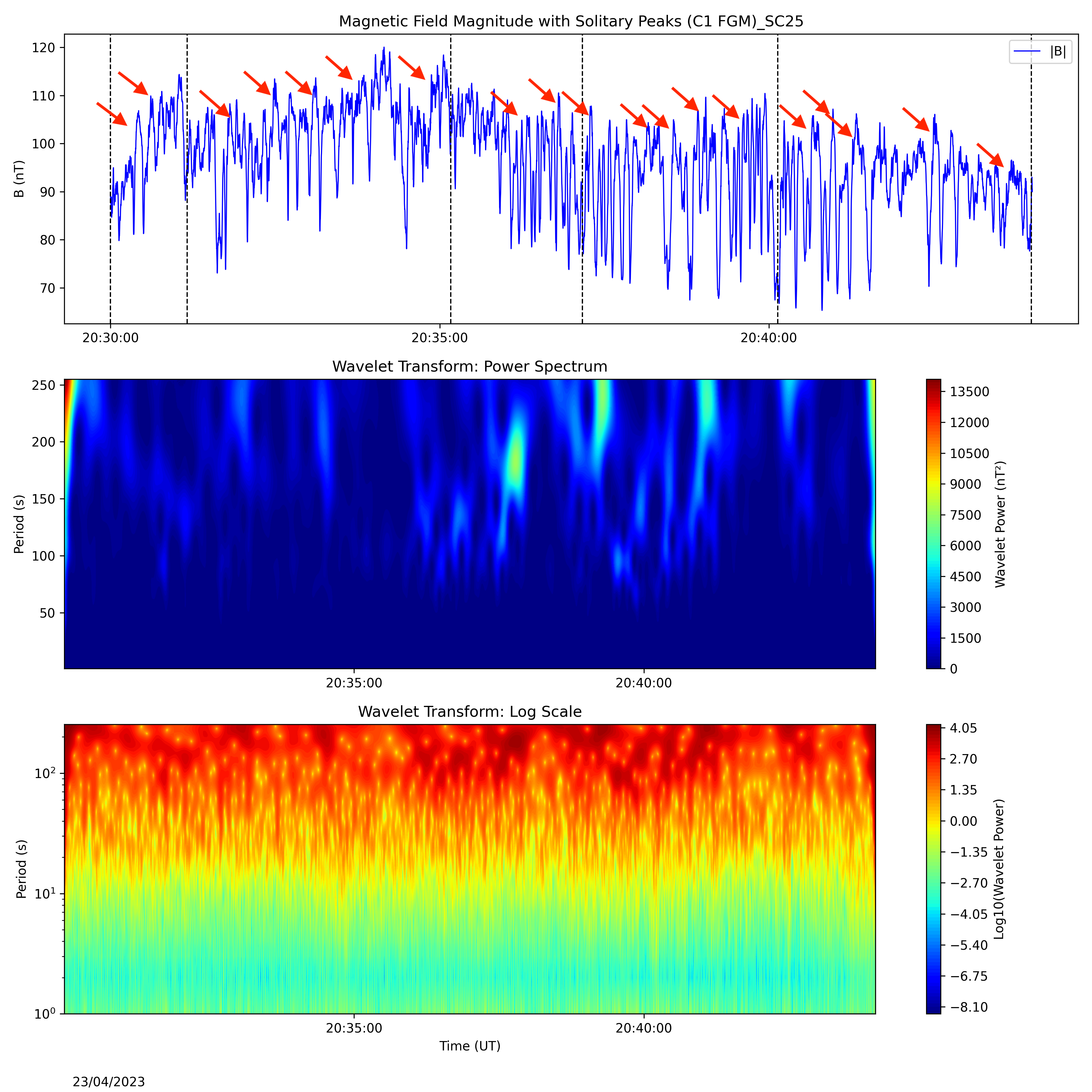}
\caption{Detection of magnetosonic solitary wave trains during the SC25 geomagnetic storm using Cluster~C1 FGM magnetic field data between 20:30 and 20:42~UT on 23~April~2023. The top panel shows the magnetic field magnitude $|B|$, where sharp localized magnetic enhancements marked by arrows represent individual soliton pulses. Vertical dashed lines separate groups of pulses, indicating multiple distinct trains of solitary structures with aperiodic amplitudes. The middle panel presents the continuous wavelet transform (CWT) power spectrum of $|B|$, highlighting temporally localized energy enhancements across characteristic periods of $\sim$50--250~s. The color scale represents wavelet power in units of nT$^2$, with higher values (yellow to red; $\sim10^4$--$1.3\times10^4$~nT$^2$) corresponding to strong localized magnetic disturbances while the bottom panel shows the logarithmic wavelet power spectrum.}
\label{fig:wavelet_analysis_SC25}
\end{figure}

\begin{figure}[H]
\includegraphics[width=\columnwidth]{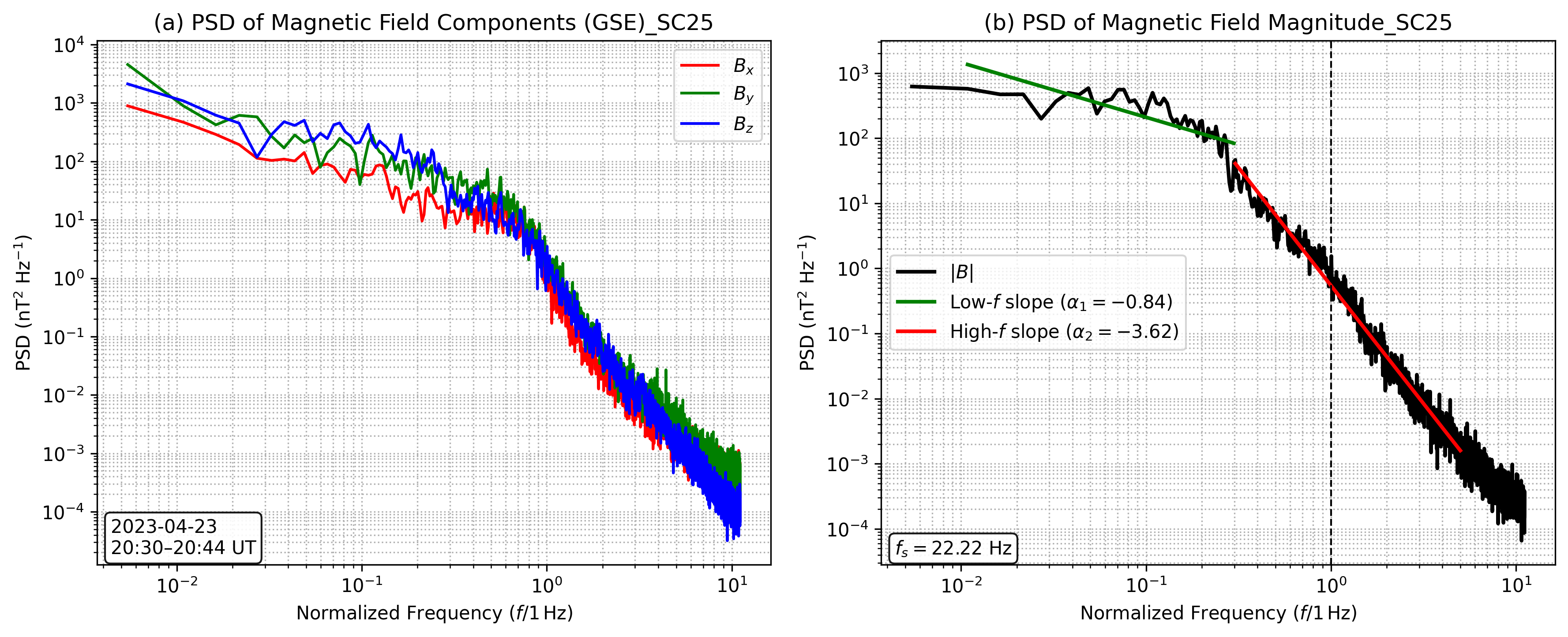}
\caption{Power spectral density (PSD) of Cluster~C1 magnetic field fluctuations during the SC25 geomagnetic storm between 20:30 and 20:44~UT on 23~April~2023. 
Panel~(a) shows the PSD of the magnetic field components ($B_x$, $B_y$, $B_z$) in GSE coordinates, illustrating broadband spectral behaviour without discrete frequency peaks. Panel~(b) shows the PSD of the magnetic field magnitude $|B|$, exhibiting a broken power-law spectrum. The vertical dashed line marks the approximate spectral break near normalized frequency $f/1\,\mathrm{Hz}\sim1$. The low-frequency range follows a shallow spectral slope $\alpha_1 \approx -0.84$, while the high-frequency range shows a steep spectral decay with slope $\alpha_2 \approx -3.62$. The absence of narrow spectral peaks and the presence of scale-dependent spectral steepening indicate nonlinear, aperiodic fluctuations consistent with magnetosonic solitary wave activity.}
\label{fig:PSD_SC25}
\end{figure}

Consistent with the SC24 observations, figure~\ref{fig:gms_solitons_SC25} shows that the SC25 FGM measurements exhibit small, localized magnetic fluctuations during the storm onset phase (20:30--20:40~UT on 23~April~2023), indicating the presence of similar early-stage nonlinear magnetic activity. The interval examined in this figure corresponds to the early phase of the storm, prior to the deepest Dst minimum, consistent with figure~\ref{fig:gms_indices_sc25}. Following the same methodology adopted for SC24, we apply the MVA, CWT, and PSD analyzes to the SC25 interval in order to characterize the temporal and scale-dependent properties of the observed magnetic fluctuations shown in figures~\ref{fig:MVA_SC25}, \ref{fig:wavelet_analysis_SC25}, and \ref{fig:PSD_SC25} respectively.

Figure~\ref{fig:MVA_SC25} shows the MVA analysis of the Cluster~C1 magnetic field over a short ($\sim$25~s) interval during the SC25 geomagnetic storm, centered on a sequence of pronounced solitary magnetic pulses. In this case, the trajectories in the hodograms show a similar pattern to that in SC24, indicating the absence of linear, monochromatic wave behavior. The compact, distorted hodogram patterns are consistent with aperiodic, nonlinear structures propagating past the spacecraft. Panel~(d) displays the magnetic field components projected onto the MVA frame where the fluctuations are dominated by the L (maximum variance) component, which exhibits sharp, intermittent excursions associated with individual solitary pulses. The M (intermediate variance) component shows weaker but organized variations, while the N (minimum variance) component remains comparatively small, indicating limited fluctuations along this direction. This ordering reflects a well-defined propagation geometry with a dominant variance axis. The eigenvalues ($\lambda_L \approx 4.0 \times 10^{1}$, $\lambda_M \approx 2.5 \times 10^{1}$, and $\lambda_N \approx 1.2 \times 10^{1}$) and their ratios ($\lambda_L/\lambda_M \approx 1.6$ and $\lambda_M/\lambda_N \approx 2.2$) demonstrate a clear separation of variance directions, confirming the coherence of the structure over the selected interval. Collectively, the enhanced pulse occurrence, the dominance of the $B_L$ component, bounded hodogram trajectories, and the absence of periodic signatures provide strong evidence that the SC25 fluctuations correspond to intensified magnetosonic solitary pulses compared to those of SC24. 

Figure~\ref{fig:wavelet_analysis_SC25} presents the wavelet analysis of the Cluster~C1 magnetic field during the early phase of the SC25 geomagnetic storm between 20:30 and 20:42~UT on 23~April~2023. Unlike SC24, where solitary structures appeared intermittently with moderate amplitudes, the SC25 interval exhibits more pronounced and densely clustered solitary pulses, indicating enhanced nonlinear wave activity. The top panel shows the magnetic field magnitude $|B|$, where sharp localized soliton peaks are marked by red arrows. The vertical dashed lines separate multiple trains of solitons, each consisting of a sequence of pulses with strongly varying amplitudes and irregular temporal spacing, showing a higher occurrence rate and stronger intermittency compared to SC24. This behavior suggests intensified episodic generation of nonlinear magnetosonic activity during SC25. The middle panel displays the continuous wavelet transform (CWT) power spectrum of $|B|$. Consistent with the SC24 case, the solitary pulses correspond to temporally localized power enhancements distributed across characteristic periods of approximately 50--250~s. However, SC25 shows substantially stronger energy concentration, with wavelet power reaching $\sim(1.0$--$1.3)\times10^{4}$~nT$^2$. The color scale represents wavelet power in units of nT$^2$, where dark blue indicates weak fluctuations ($<10^3$~nT$^2$), green–yellow represents moderate energy levels, and bright yellow–red regions indicate intense localized magnetic disturbances associated with strong solitary pulses. Similar to SC24, the absence of persistent horizontal spectral bands confirms the non-stationary and aperiodic nature of the fluctuations. The bottom panel shows the logarithmic wavelet power spectrum, highlighting broadband energy distributions extending across multiple temporal scales. The vertically elongated power structures remain temporally confined but appear more frequent and intense than in SC24, indicating stronger scale coupling and enhanced nonlinear dispersive interactions. 

Overall, the increased number of pulses, stronger wavelet power, and the presence of multiple clustered soliton trains demonstrate that nonlinear magnetosonic solitary activity is significantly more pronounced during SC25 than SC24. These observations suggest that the storm-time plasma environment during SC25 supported stronger nonlinear wave growth and intermittent energy localization.

Similar to the SC24 interval, the PSD of the Cluster~C1 magnetic field during SC25 also exhibits broadband spectral behavior indicative of nonlinear, aperiodic fluctuations. As shown in Figure~\ref{fig:PSD_SC25}(a), all three magnetic field components display smooth, continuous spectra without discrete spectral peaks, confirming the absence of coherent monochromatic wave activity. The PSD of the magnetic field magnitude $|B|$ in Figure~\ref{fig:PSD_SC25}(b) again reveals a broken power-law structure, with a spectral transition occurring near the normalized frequency $f/1\,\mathrm{Hz}\sim1$. Below the spectral break, the fluctuations follow a relatively shallow scaling with a slope $\alpha_1 \approx -0.84$, indicating dominant large-scale compressive dynamics. Above the break, the spectrum steepens significantly to $\alpha_2 \approx -3.62$, reflecting enhanced dispersive and kinetic-scale nonlinear processes. Compared to SC24, the SC25 spectrum exhibits a noticeably flatter low-frequency slope and a steeper high-frequency decay, suggesting stronger intermittency and more pronounced nonlinear steepening during this interval. Consistent with the SC24 observations, the absence of narrow spectral peaks and the presence of scale-dependent spectral steepening confirm that the fluctuations are not associated with linear periodic waves but instead correspond to aperiodic magnetosonic solitary structures. These spectral characteristics further support the interpretation that nonlinear wave activity is intensified during the SC25 event.

\begin{figure}[H]
  \centering
  \includegraphics[width=\columnwidth]{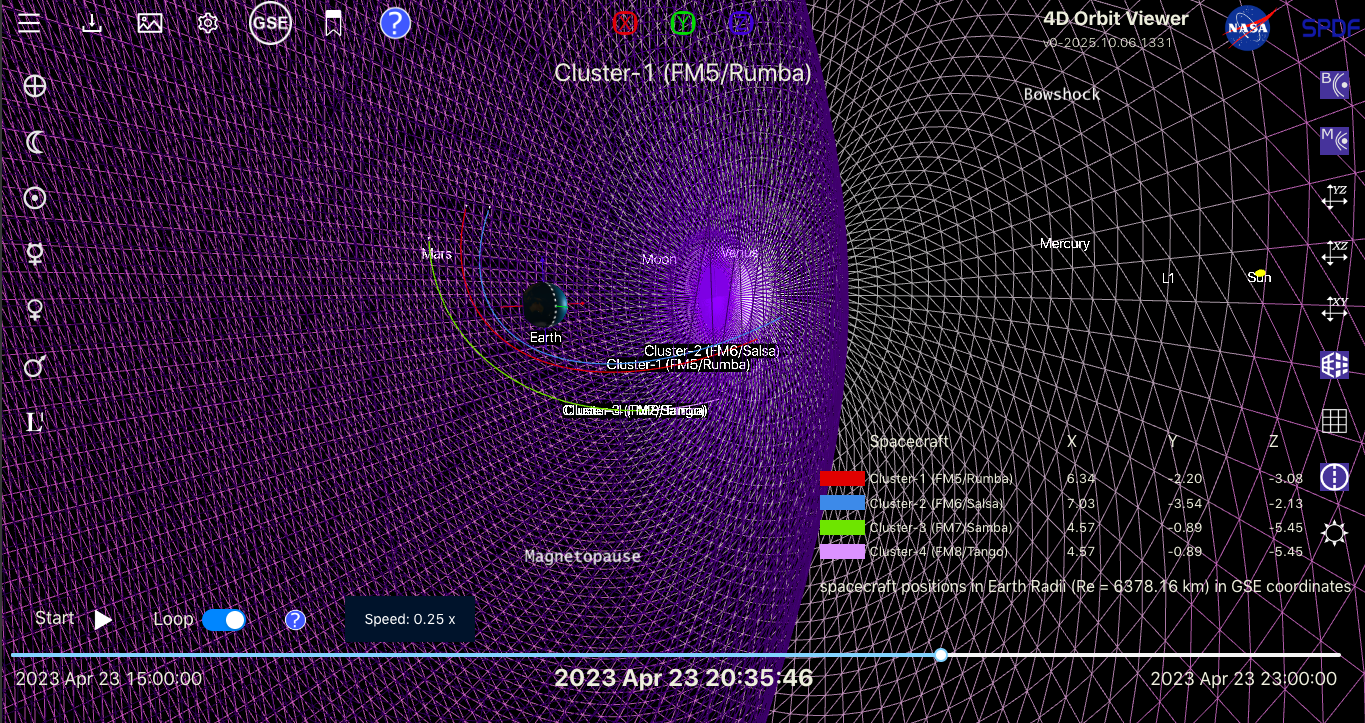}\\
  \caption{Orbital configuration of the Cluster constellation in GSE coordinates during the SC25 geomagnetic storm on 23~April~2023, shown using the 4D Orbit Viewer \citep{ssc4dviewer}. The trajectories of the four Cluster spacecrafts are plotted relative to the modeled bow shock and magnetopause. The snapshot corresponds to $\sim$20:35:46~UT, within the interval 20:30:00--20:40:00~UT during which magnetosonic solitary structures were detected by Cluster--1 as it crossed the magnetopause region of the magnetosphere.}
  \label{fig: spacecraft_location_SC25}
\end{figure}

Figure~\ref{fig: spacecraft_location_SC25} shows the orbital configuration of the Cluster spacecraft in GSE coordinates during the SC25 geomagnetic storm on 23~April~2023. The snapshot at $\sim$20:35:46~UT corresponds to the interval when solitary waves or pulses were detected by Cluster~C1. During this time, the spacecraft was located in the dayside magnetospheric boundary region, close to the magnetopause and well inside the bow shock. A comparison with the SC24 event shows that solitary structures were detected under similar orbital conditions, with Cluster~C1 traversing the magnetopause boundary region in both storms. This recurrence suggests that the magnetopause provides favorable plasma conditions, including FLR effects and kinetic-scale processes, for the generation of nonlinear magnetosonic structures \citep{Bora2019, Stasiewicz1993}. The localized spacecraft sampling further supports that the observed pulses represent spatially confined magnetic structures rather than global disturbances.

\begin{figure}[htbp]
  \centering
  \includegraphics[width=\columnwidth]{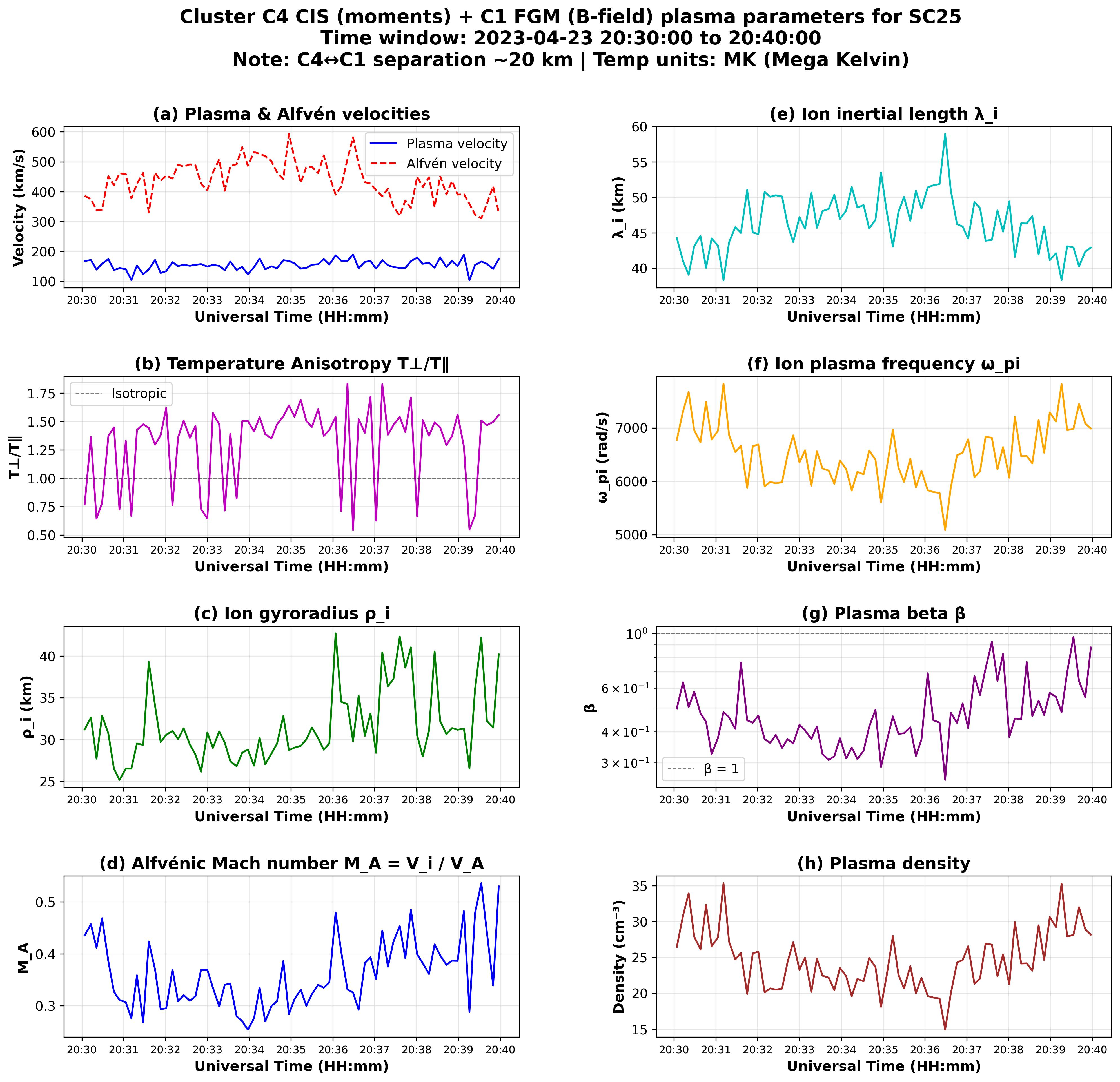}\\
  \caption{The plasma and magnetic field parameters during the interval 20:30--20:40~UT on 23~April~2023 (SC25), derived from the Cluster C4 CIS ion moments and C1 FGM magnetic field measurements. Panels~(a)--(d) display the plasma bulk velocity, Alfv\'en velocity, temperature anisotropy $T_{\perp}/T_{\parallel}$, and Alfv\'enic Mach number $M_A$, respectively, while panels~(e)--(h) present the ion inertial length $\lambda_i$, ion plasma frequency $\omega_{pi}$, plasma beta $\beta$, and plasma density. The C4--C1 spacecraft separation during this interval is $\sim$20~km, ensuring near-coincident sampling of plasma and magnetic field conditions within the same localized region.}
  \label{fig:plasma_params_SC25}
\end{figure}

Figure~\ref{fig:plasma_params_SC25} presents the plasma and magnetic field–derived parameters during the early phase of the SC25 geomagnetic storm between 20:30 and 20:40~UT on 23~April~2023. The plasma flow velocity remains consistently lower than the local Alfv\'en speed ($V_i \approx 100$--$180$~km~s$^{-1}$ and $V_A \approx 310$--$600$~km~s$^{-1}$), resulting in sub-Alfv\'enic flow conditions with $M_A \approx 0.22$--$0.58$. Such flow regimes favor the propagation and stability of nonlinear compressive magnetosonic structures within magnetically dominated plasma environments. The ion temperature anisotropy exhibits persistent deviations from isotropy, with $T_{\perp}/T_{\parallel}$ varying between $\sim0.5$ and 1.8 and frequently exceeding unity, indicating sustained perpendicular ion heating and the presence of free energy capable of driving nonlinear compressive modes. The ion gyroradius $\rho_i$ varies between $\sim25$ and 45~km, while the ion inertial length $\lambda_i$ remains within $\sim35$--60~km, closely matching the characteristic spatial scales of the observed magnetic pulses. This correspondence suggests that finite Larmor radius and ion inertial effects play a dominant role in controlling the dispersive properties responsible for MS soliton formation. The ion plasma frequency remains relatively stable at $\omega_{pi} \sim (5.2$--$7.8)\times10^{3}$~rad~s$^{-1}$, while the plasma density fluctuates between $\sim15$ and 36~cm$^{-3}$, indicating moderate density compression. The plasma beta remains below or close to unity ($\beta \approx 0.25$--0.9) throughout most of the interval, implying magnetically dominated conditions favorable for pressure-balanced nonlinear magnetic structures. The overall parameter ranges are summarized in Table~\ref{fig:Table2_SC25}.

\begin{table*}[htb]
\centering
\caption{Plasma parameters observed during the April 2023 (SC25) }
\label{table:plasma_params}
\begin{tabular}{lp{5cm}p{4cm}p{7cm}}
\toprule
\textbf{S.no} & \textbf{Parameter} & \textbf{Value Range} \\
\midrule
1 & Plasma velocity & 100-180 km/s \\
2 & Alfvén velocity & 310-600 km/s \\
3 & Temperature anisotropy & 0.5-1.8 \\ 
4 & Ion gyroradius & 25-45 km \\
5 & Alfvénic Mach no. & 0.22-0.58 \\
6 & Ion inertial length & 35-60 km \\
7 & Ion plasma frequency & 5200-7800 rad/s\\
8 & Plasma beta & 0.25-0.9 \\
9 & Plasma density & 15-36 cm$^{-3}$ \\
\bottomrule
\end{tabular}
\label{fig:Table2_SC25}
\end{table*}

\vspace{0em}
\section{Conclusion}
This study presents a comparative investigation of nonlinear magnetosonic (MS) solitary waves observed during geomagnetic storms across Solar Cycles 24 and 25 using high-resolution multipoint magnetic field measurements from the Cluster II mission. We analyse two intense geomagnetic storm events, one from each solar cycle (March 2015 for SC24 and April 2023 for SC25), selected based on clear and prominent signatures of MS solitary wave activity, characterised by localized magnetic field fluctuations resembling MS solitons \citep{Stasiewicz2003, Bora2019}. The SC24 event corresponds to the well-known St. Patrick’s Day storm of March 2015, one of the most severe space weather events of Solar Cycle 24. This storm was primarily driven by an Earth-directed coronal mass ejection (CME) launched on 15 March 2015, whose interplanetary shock impacted Earth on 17 March, producing strong geomagnetic disturbances with minimum Dst values approaching −223 nT and Kp levels reaching G4 storm intensity. The SC25 event corresponds to the intense geomagnetic storm of 23–24 April 2023, triggered by a CME originating from a relatively modest active region but producing unexpectedly strong geomagnetic activity. The SC25 storm reached comparable G4 severity, with Dst decreasing below −230 nT and Kp values approaching 8, indicating enhanced solar wind–magnetosphere coupling efficiency. Despite differences in solar source characteristics, both storms generated strong magnetospheric compression and boundary layer restructuring, creating favourable plasma environments for nonlinear magnetosonic solitary wave formation. The comparison of these two major geomagnetic storms, therefore, provides insight into the evolution of solitary wave activity under different solar cycle conditions within similar magnetopause boundary regions.

Minimum variance analysis (MVA), widely employed in earlier studies \citep{Stasiewicz2003, Bora2019}, provides a fundamental diagnostic for determining the geometry of MS solitary structures. However, MVA alone may be insufficient when magnetic fluctuations are weak, turbulence-embedded or inadequately resolved due to limited temporal cadence. Since the detection of MS solitary waves strongly depends on sampling resolution, which is often unavailable for several space missions, we complement MVA with additional advanced diagnostics. These include continuous wavelet transform (CWT) analysis for multiscale time–frequency localization, hodogram analysis for examining magnetic field rotation, power spectral density (PSD) analysis for assessing nonlinear spectral behavior, and spacecraft localization to establish environmental context. The integration of these complementary diagnostics provides a robust, reproducible, and mission-independent observational framework for reliably identifying and characterizing MS solitary wave trains in space plasmas.

The results demonstrate that localized MS solitary structures occur during the early storm intervals in both SC24 and SC25, suggesting that nonlinear wave activity represents a characteristic magnetospheric response to enhanced solar wind forcing prior to the full development of geomagnetic storm main phases. During SC24, solitary structures were detected near the magnetopause boundary under trans-Alfvénic flow conditions, moderate plasma beta, and ion kinetic scale lengths comparable to the spatial extent of the observed magnetic pulses, indicating that finite Larmor radius (FLR) and dispersive effects were already influencing the nonlinear wave dynamics. In contrast, solitary wave activity during SC25 is considerably stronger and more frequent. The enhanced soliton activity observed during SC25 is likely associated with stronger solar wind compression, increased plasma density variability, enhanced temperature anisotropy fluctuations, and comparatively larger ion kinetic scale lengths. These conditions intensify FLR-mediated dispersive effects and promote nonlinear steepening, resulting in the formation of more coherent and densely packed MS soliton trains. Although both events occurred near the magnetopause boundary, the plasma environment during SC25 appears to have supported stronger nonlinear energy localization due to intensified solar wind–magnetosphere coupling. These findings indicate that MS solitons may represent diagnostic signatures of evolving geomagnetic storm activity and reflect localized nonlinear energy transfer and boundary layer reconfiguration processes. While further statistical investigations across multiple storm events are necessary to establish predictive capability, the present results suggest their potential role as precursor-like indicators of magnetospheric disturbance.

A clear contrast in solitary wave behavior is identified between the two solar cycles. While MS solitary structures are consistently observed during the early storm phases in both cycles, confirming that the magnetopause boundary region supports nonlinear dispersive processes, the SC24 event exhibits comparatively weaker and more turbulence-embedded solitary structures. In contrast, the SC25 event reveals stronger, denser, and more organized soliton trains characterized by enhanced wavelet power, steeper high-frequency spectral decay, and stronger eigenvalue separation in the MVA frame.

The multi-diagnostic framework developed in this study is adaptable and can be applied to single-spacecraft missions lacking multipoint measurements. The combined use of MVA coordinate transformation, wavelet-based time–frequency localization, broadband spectral analysis, and plasma kinetic scale diagnostics provides a mission-independent methodology for identifying nonlinear solitary wave activity, even with a lower cadence. This approach is particularly relevant for solar and heliospheric missions such as Wind, Solar Orbiter, Parker Solar Probe, and Aditya-L1, where identifying localized nonlinear plasma structures is essential for understanding solar wind turbulence, particle acceleration, and energy transport in collisionless plasmas. Compared to magnetospheric missions like Cluster II, which often provide high-cadence measurements, several heliospheric missions operate with relatively lower temporal resolution, making the detection of short-duration solitary structures challenging. The methodology presented here mitigates this limitation by integrating complementary diagnostics that allow solitary structures to be identified even when individual pulses are partially resolved or embedded within background fluctuations. Application of this framework to solar wind and coronal plasma observations can, therefore, significantly improve the detection of nonlinear wave activity and enhance the interpretation of in-situ solar plasma measurements.

Future work should extend this comparative approach across a broader range of geomagnetic storms and solar cycle conditions and incorporate observations from MMS, THEMIS, and heliospheric spacecraft. Such investigations will help establish statistical occurrence patterns of MS solitons, clarify their role in magnetospheric energy dissipation, and evaluate their reliability as indicators of space weather dynamics.

\section*{Acknowledgements}
The authors express their sincere gratitude to the Cluster Science Archive (CSA) operated by the European Space Agency (ESA) and the NASA CDAWeb database for providing access to the Cluster II mission magnetic field and plasma data used throughout this study. We also acknowledge the World Data Center for Geomagnetism, Kyoto (WDC Kyoto), and  Helmholtz Center for Geosciences (GFZ) for supplying the Dst and Kp geomagnetic indices, which were essential for characterizing the storm-time conditions of the events analyzed. We are grateful to NASA’s 4D Orbit Viewer tool for enabling precise spacecraft position visualization and for supporting the interpretation of the Cluster spacecraft trajectories during the SC24 and SC25 intervals. The availability of these high-quality datasets and visualization resources made this work possible. We also extend our appreciation to all organizations maintaining open-access scientific platforms for the advancement of space physics research. The author, Yimnasangla, gratefully acknowledges the University of Science and Technology Meghalaya (USTM), Meghalaya, India, for the internship opportunity during which this work was initiated. The author (M.K.) gratefully acknowledges support from the ongoing ISRO Research Project (Proposal No.~RAC-S/GU/2024/4/51), under which this work was carried out as part of the ISRO RAC-S@GU program.
\vspace{-1em}

\bibliographystyle{elsarticle-harv}
\bibliography{cas-refs}

\end{document}